\documentclass{emulateapj}
\usepackage{graphicx}
\usepackage{multirow,natbib}
\usepackage{textcomp}
\usepackage{epsfig}
\usepackage{amsmath}
\usepackage{amssymb}
\usepackage{amsthm}
\usepackage{xy}

\def\uu {4U\,0142$+$614}
\def\ee {1E\,1048.1$-$5937}

\def\xte{XTE\,J1810$-$197\,}

\def\wes{CXOU\,J1647$-$4552}

\def\sgre{SGR\,0501$+$4516}
\def\sgrf{SGR\,0418$+$5729}

\def\psr{PSR\,1622$-$4950}
\def\hbpsr{PSR\,J1846$-$0258}

\newcommand{\chandra}{{\it Chandra}}
\newcommand{\rosat}{{\em ROSAT}}

\newcommand{\xmm}{{\em XMM--Newton}}

\newcommand{\suzaku}{{\em Suzaku}}
\newcommand{\rxte}{{\em RXTE}}
\newcommand{\swift}{{\em Swift}}
\newcommand{\fermi}{{\em Fermi}}

\newcommand{\bc}{\begin{center}}
\newcommand{\ec}{\end{center}}
\def\ltsima{$\; \buildrel < \over \sim \;$}
\def\lsim{\lower.5ex\hbox{\ltsima}}
\def\loe{\lower.5ex\hbox{\ltsima}}
\def\gtsima{$\; \buildrel > \over \sim \;$}
\def\gsim{\lower.5ex\hbox{\gtsima}}
\def\goe{\lower.5ex\hbox{\gtsima}}

\def\ltsima{$\; \buildrel < \over \sim \;$}
\def\lsim{\lower.5ex\hbox{\ltsima}}
\def\loe{\lower.5ex\hbox{\ltsima}}
\def\gtsima{$\; \buildrel > \over \sim \;$}
\def\gsim{\lower.5ex\hbox{\gtsima}}
\def\goe{\lower.5ex\hbox{\gtsima}}
\def\nh {$N_{H}$}

\def\ergs {erg\,s$^{-1}$}
\def\ergscm2 {erg\,s$^{-1}$cm$^{-2}$}

\def\cm2 {cm$^{-2}$}
\def\arcsec{$^{\prime\prime}$}
\def\arcmin{$^{\prime}$}

\def\ergs {${\rm erg\, s}^{-1}$}

\def\src {Swift\,J1822.3--1606}
\def\flux {\mbox{erg cm$^{-2}$ s$^{-1}$}}

\def\nh {$N_{\rm H}$}

\shorttitle{A NEW LOW-B MAGNETAR:  Swift\,J1822.3--1606}
\shortauthors{N. REA ET AL.}

\begin{document}

\title{A new low magnetic field magnetar: the 2011 outburst of Swift\,J1822.3--1606}

\author{N. Rea\altaffilmark{1}, G. L. Israel\altaffilmark{2}, P. Esposito\altaffilmark{3}, J. A. Pons\altaffilmark{4}, A. Camero-Arranz\altaffilmark{1}, 
R. P. Mignani\altaffilmark{5,6}, R. Turolla\altaffilmark{7,5}, S. Zane\altaffilmark{5}, 
M. Burgay\altaffilmark{3}, A. Possenti\altaffilmark{3},  
S. Campana\altaffilmark{8},  
T. Enoto\altaffilmark{9}, 
N. Gehrels\altaffilmark{10}, E. G{\"o}{\u g}{\"u}{\c s}\altaffilmark{11}, D. G\"otz\altaffilmark{12}, C. Kouveliotou\altaffilmark{13},
K. Makishima\altaffilmark{14,15}, 
S. Mereghetti\altaffilmark{16}, S. R. Oates\altaffilmark{5}, 
D. M. Palmer\altaffilmark{17} 
R. Perna\altaffilmark{18}, L. Stella\altaffilmark{2}, A. Tiengo\altaffilmark{19,16}
}

\affil{$^1$ Institut de Ci\`encies de l'Espai (CSIC-IEEC), Campus UAB, Facultat de Ci\`encies, 
Torre C5-parell, E-08193 Barcelona, Spain}
\affil{$^2$ INAF -- Osservatorio Astronomico di Roma, via Frascati 22, 00040, Monteporzio Catone, Italy}
\affil{$^3$ INAF -- Osservatorio Astronomico di Cagliari, localit\`a Poggio dei Pini, strada 54, I-09012 Capoterra, Italy}
\affil{$^4$ Departament de Fisica Aplicada, Universitat d'Alacant, Ap. Correus 99, 03080 Alacant, Spain}
\affil{$^5$ Mullard Space Science Laboratory, University College London, Holmbury St. Mary, Dorking, Surrey RH5 6NT, UK}
\affil{$^6$ Institute of Astronomy, University of Zielona G\'ora, Lubuska 2, 65-265, Zielona G\'ora, Poland}
\affil{$^7$ Universit\`a di Padova, Dipartimento di Fisica e Astronomia, via F.~Marzolo 8, I-35131 Padova, Italy}
\affil{$^8$ INAF -- Osservatorio Astronomico di Brera, via E.~Bianchi 46, I-23807 Merate, Italy}
\affil{$^9$ Kavli Institute for Particle Astrophysics \& Cosmology (KIPAC), SLAC/Stanford University, PO Box 20450, MS 29, Stanford, CA 94309, USA}
\affil{$^{10}$ NASA Goddard Space Flight Center, Greenbelt, MD 20771, USA}
\affil{$^{11}$ Sabanc\i\ University, Orhanl\i-Tuzla, 34956 \.Istanbul, Turkey}
\affil{$^{12}$ AIM (CEA/DSM-CNRS-Universit\'e Paris Diderot), Irfu/Service d'Astrophysique, Saclay, F-91191 Gif-sur-Yvette, France}
\affil{$^{13}$ NASA Marshall Space Flight Center, Huntsville, AL 35812, USA}
\affil{$^{14}$ High Energy Astrophysics Laboratory, Institute of Physical and Chemical Research (RIKEN), Wako, Saitama 351-0198, Japan}
\affil{$^{15}$ Department of Physics, University of Tokyo, 7-3-1 Hongo, Bunkyo-ku, Tokyo 113-0033, Japan}
\affil{$^{16}$ INAF -- IASF Milano, via E. Bassini 15, I-20133 Milano, Italy}
\affil{$^{17}$ Los Alamos National Laboratory, Los Alamos, NM 87545, USA}
\affil{$^{18}$ JILA, University of Colorado, Boulder, CO 80309-0440, USA}
\affil{$^{19}$ IUSS Ð Istituto Universitario di Studi Superiori, viale Lungo Ticino Sforza 56, 27100 Pavia, Italy}


\begin{abstract}

We report on the long term X-ray monitoring with \swift, \rxte,
\suzaku, \chandra, and \xmm\ of the outburst of the newly discovered magnetar
\src\,(SGR\,1822-1606), from the first observations soon after the
detection of the short X-ray bursts which led to its discovery,
through the first stages of its outburst decay (covering the time-span
from July 2011, until end of April  2012). We also report on
archival \rosat\ observations which witnessed the source during its
likely quiescent state, and on upper limits on \src's radio-pulsed and
optical emission during outburst, with the Green Bank Telescope (GBT)
and the Gran Telescopio Canarias (GTC), respectively. Our X-ray timing
analysis finds the source rotating with a period of
$P=8.43772016(2)$\,s and a period derivative
$\dot{P}=8.3(2)\times10^{-14}$~s\,s$^{-1}$ , which entails an inferred dipolar
surface magnetic field of $B\simeq2.7\times10^{13}$~G at the equator. This
measurement makes \src\, the second lowest magnetic field magnetar (after \sgrf; Rea et al. 2010). Following
the flux and spectral evolution from the beginning of the outburst, 
we find that the flux decreased by about an order of
magnitude, with a subtle softening of the spectrum, both typical of
the outburst decay of magnetars. By modeling the secular thermal
evolution of \src, we find that the observed timing properties of the
source, as well as its quiescent X-ray luminosity, can be reproduced if it
was born with a poloidal and crustal toroidal fields of
$B_{p}\sim1.5\times10^{14}$\, G and $B_{tor}\sim7\times10^{14}$\, G,
respectively, and if its current age is $\sim$550\,kyr.

\end{abstract}

\section{Introduction}\label{intro}

The availability of sensitive, large field-of-view X-ray monitors such
as the Burst Alert Telescope on board \swift, and the Gamma-ray Burst
Monitor on \fermi, makes us witness a golden age for magnetar
studies. Since the discovery of the first magnetar outbursts \citep{gavriil02,kouveliotou03,kaspi03,ibrahim04}, five new members of
the class have been discovered through the serendipitous detection of
the typical short X-ray bursts emitted by these highly energetic X-ray
pulsars, and the accompanying increase in the persistent emission (see
\citealt{rea11} for a recent review on magnetar outbursts).

Magnetars, usually recognized in the anomalous X-ray pulsar (AXP) and
soft gamma-ray repeater (SGR) classes, are isolated neutron stars with
bright persistent X-ray emission ($L_{\rm X}\sim10^{33}$--$10^{36}$
\ergs ), rotating at spin periods of $\sim$0.3--12~s and with large
period derivatives ($10^{-13}$--$10^{-10}$ s\,s$^{-1}$; see
\citealt{mereghetti08, rea11} for a review). Sporadically, they emit
bursts and flares which can last from a fraction of seconds to minutes,
releasing $\sim$$10^{38}$--$10^{47}$ \ergs , and are often accompanied
by long-lived (up to years) increases of the persistent X-ray
luminosity (outbursts).

\begin{figure*}
\hspace{-1cm}
\includegraphics[height=8cm]{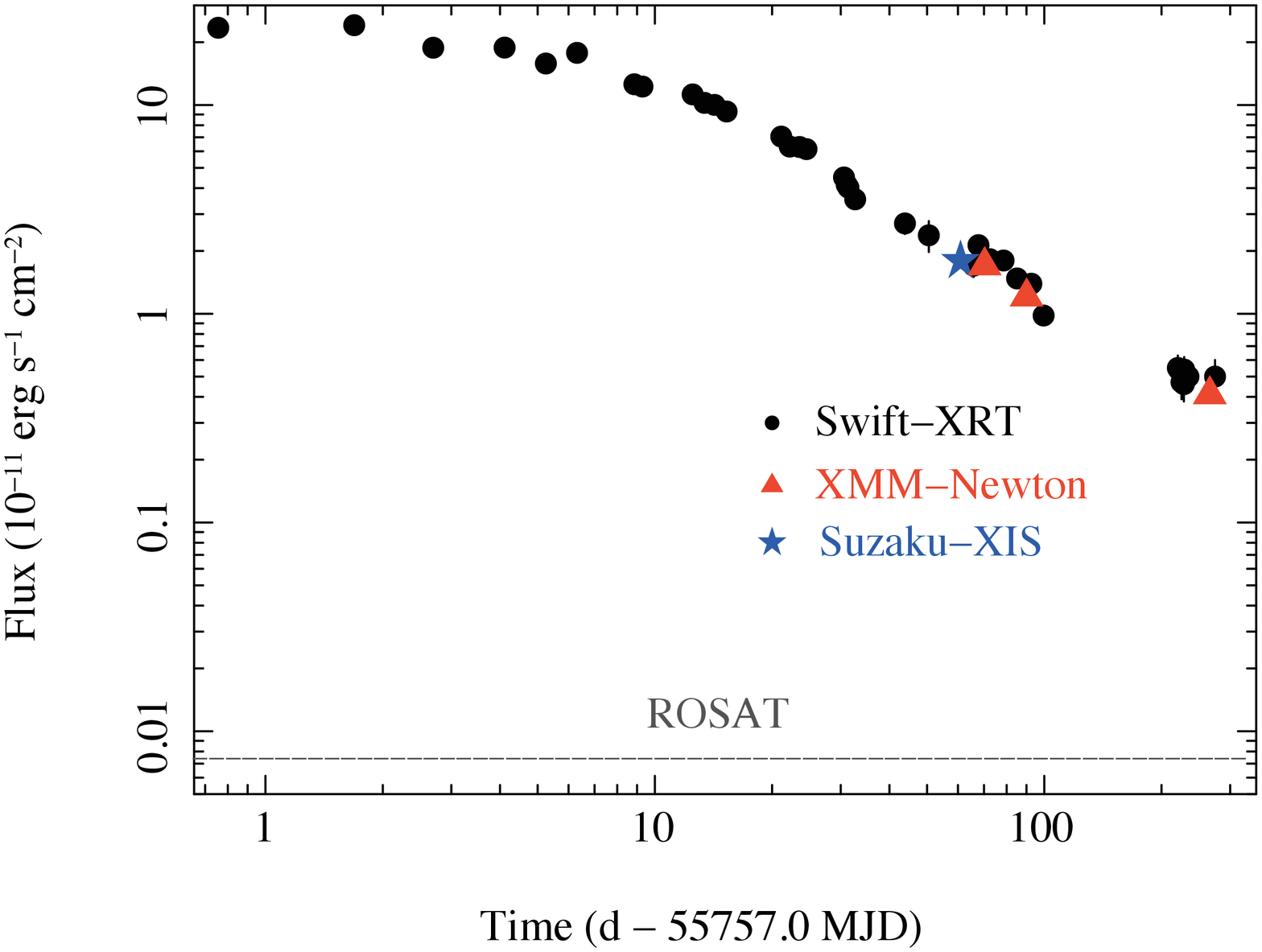}
\hspace{-1cm}
\includegraphics[height=10cm]{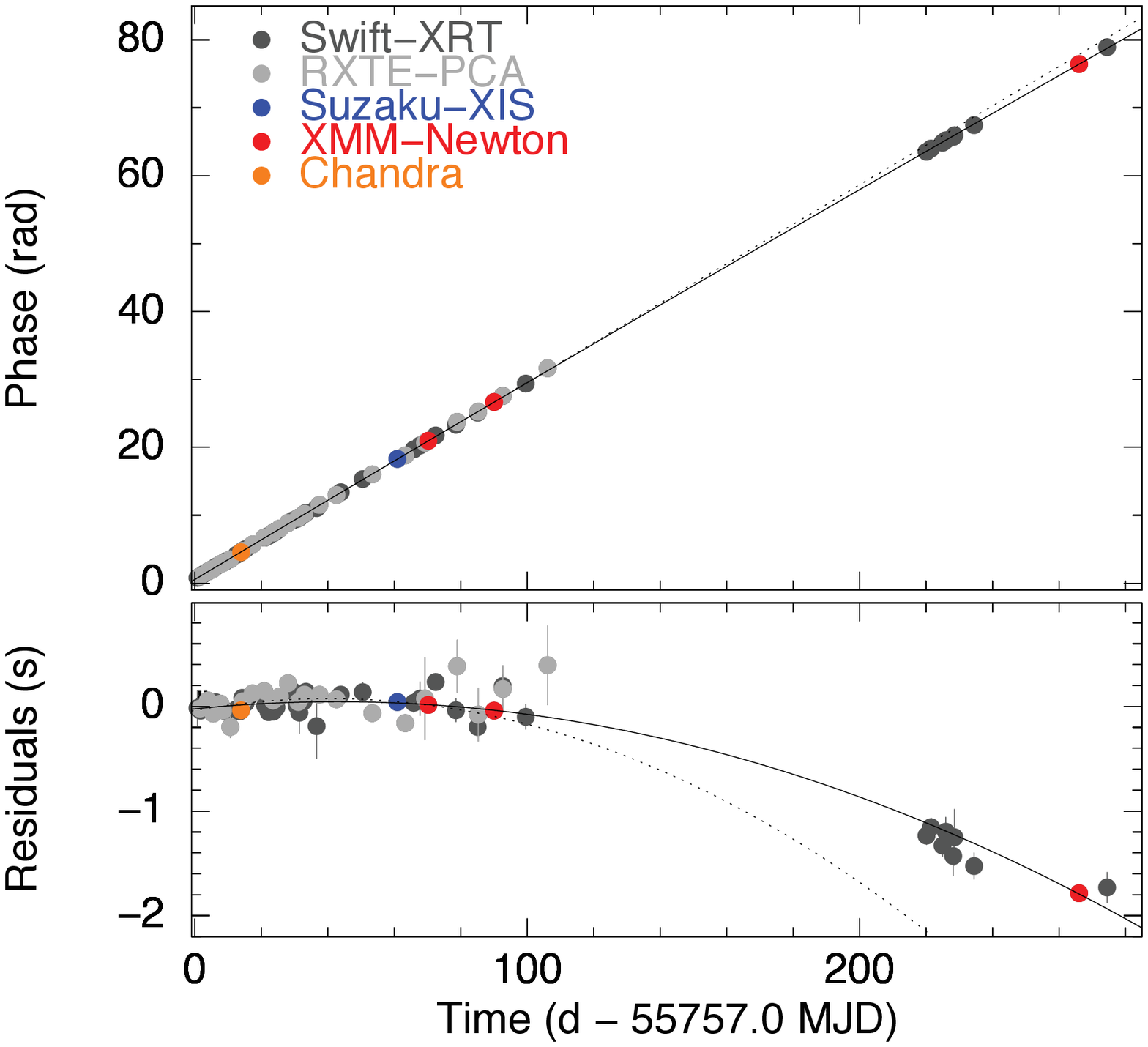}
\caption{{\em Left panel}: Flux decay of \src\, in the 1--10\,keV energy range (see also Table\,\ref{tab:spec}). {\em Right panel}: pulse phase evolution as a function of time, together with the time residuals (lower panel) after having corrected for the linear component (correction to the $P$ value). The solid lines in the two panels mark the inferred  $P$--$\dot{P}$ coherent solution based on the whole dataset, while the dotted lines represent the $P$--$\dot{P}$ coherent solution based on the data collected during the first 90 days only  (see text for the details). }
\label{fig:phases}
\end{figure*}

\begin{table}
\centering
\caption{\src\, observations used for this work.}
\label{obs-log}
{\footnotesize
\begin{tabular}{@{}cccc}
\hline
\hline
Instrument & Obs.ID & Date$^{a}$ & Exposure\\
 & & (MJD TBD) & (ks)\\
\hline
\swift/XRT & 00032033001 (PC) & 55\,757.75058 & 1.6 \\
\rxte/PCA  & 96048-02-01-00   & 55\,758.48165& 6.5 \\
\swift/XRT & 00032033002 (WT) & 55\,758.68430 & 2.0 \\
\swift/XRT & 00032033003 (WT) & 55\,759.69082 & 2.0 \\
\rxte/PCA  & 96048-02-01-05   & 55\,760.80853 & 1.7 \\
\swift/XRT & 00032033005 (WT) & 55\,761.54065 & 0.5 \\
\rxte/PCA  & 96048-02-01-01   & 55\,761.55969 & 5.0 \\
\swift/XRT & 00032033006 (WT) & 55\,762.24089 & 1.8 \\
\rxte/PCA  & 96048-02-01-02   & 55\,762.47384 & 4.9 \\
\swift/XRT & 00032033007 (WT) & 55\,763.30400 & 1.6 \\
\rxte/PCA  & 96048-02-02-00   & 55\,764.61846 & 6.1 \\
\rxte/PCA  & 96048-02-02-01   & 55\,765.46687 & 6.8 \\
\swift/XRT & 00032033008 (WT) & 55\,765.85252 & 2.2 \\
\swift/XRT & 00032033009 (WT) & 55\,766.28340 & 1.7 \\
\rxte/PCA  & 96048-02-02-02   & 55\,767.59064 & 3.0 \\
\rxte/PCA  & 96048-02-02-03   & 55\,769.35052 & 3.4 \\
\swift/XRT & 00032033010 (WT) & 55\,769.49531 & 2.1 \\
\swift/XRT & 00032033011 (WT) & 55\,770.39936 & 2.1 \\
\chandra/HRC-I & 13511 & 55\,770.83049 & 11.7 \\
\swift/XRT & 00032033012 (WT) & 55\,771.23302 & 2.1 \\
\rxte/PCA  & 96048-02-03-00   & 55\,771.34185 & 6.8 \\
\swift/XRT & 00032033013 (WT) & 55\,772.40044 & 2.1 \\
\rxte/PCA  & 96048-02-03-01   & 55\,774.34999 & 6.9 \\
\rxte/PCA  & 96048-02-03-02   &  55\,777.85040 & 1.9 \\
\swift/XRT & 00032051001 (WT) & 55\,778.10744 & 1.7 \\
\swift/XRT & 00032051002 (WT) & 55\,779.18571 & 1.7 \\
\rxte/PCA  & 96048-02-04-00   &  55\,780.85040 & 6.7 \\
\swift/XRT & 00032051003 (WT) & 55\,780.49505 & 2.3 \\
\swift/XRT & 00032051004 (WT) & 55\,781.49878 & 2.3 \\
\rxte/PCA  & 96048-02-04-01   &  55\,782.57749 &  6.2\\
\rxte/PCA  & 96048-02-04-02   &  55\,784.97179 &  6.2\\
\swift/XRT & 00032051005 (WT) & 55\,786.42055 & 2.2 \\
\swift/XRT & 00032051006 (WT) & 55\,787.58688  & 2.2 \\
\rxte/PCA  & 96048-02-05-00   &  55\,788.05419 &  6.0\\
\swift/XRT & 00032051007 (WT) & 55\,788.25617  & 2.3 \\
\swift/XRT & 00032051008 (WT) & 55\,789.66173  & 1.7 \\
\rxte/PCA  & 96048-02-05-01   &  55\,789.95880 &  6.0\\
\swift/XRT & 00032051009 (WT) & 55\,790.36270  & 2.2 \\
\rxte/PCA  & 96048-02-06-00   &  55\,794.45899 &  6.5\\
\rxte/PCA  & 96048-02-07-00   &  55\,799.61550 &  6.9\\
\swift/XRT & 00032033015 (WT) & 55\,800.86278  & 2.9\\
\swift/XRT & 00032033016 (WT) & 55\,807.48660  & 2.4\\
\rxte/PCA  & 96048-02-08-00   &  55\,810.37979 &  6.0\\
\suzaku/XIS & 906002010 & 55\,817.92550 & 33.5 \\
\rxte/PCA  & 96048-02-10-00   &  55\,820.23970 &  6.7\\
\swift/XRT & 00032033017 (WT) & 55\,822.82836  & 4.9\\
\swift/XRT & 00032033018 (WT) & 55\,824.71484  & 1.5\\
\rxte/PCA  & 96048-02-10-01   & 55826.18540 & 5.6 \\
\xmm & 0672281801 & 55\,827.25350 & 10.6 \\
\swift/XRT & 00032033019 (WT) & 55\,829.45421  & 2.3\\
\swift/XRT & 00032033020 (WT) & 55\,835.54036  & 2.6\\
\rxte/PCA  & 96048-02-11-00   & 55835.90370 & 7.0 \\
\swift/XRT & 00032033021 (WT) & 55\,842.06040  & 4.2\\
\rxte/PCA  & 96048-02-12-00   & 55842.23269 & 5.8\\
\xmm & 0672282701 &  55\,847.06380 & 25.8 \\
\swift/XRT & 00032033022 (WT) & 55\,849.61916  & 3.4\\
\rxte/PCA  & 96048-02-13-00   & 55849.6597976 & 5.6 \\ 
\swift/XRT & 00032033024 (WT) & 55\,862.59155  & 10.2\\
\rxte/PCA  & 96048-02-14-00   & 55863.11100 & 5.6 \\ 
\swift/XRT & 00032033025 (PC) & 55\,977.16600& 6.3 \\
\swift/XRT & 00032033026 (WT) & 55\,978.53399 & 10.2 \\
\swift/XRT & 00032033027 (PC) & 55\,981.99499  & 11.0 \\
\swift/XRT & 00032033028 (WT) & 55\,982.96299   & 7.0\\
\swift/XRT & 00032033029 (WT) & 55\, 985.17799 & 7.0 \\
\swift/XRT & 00032033030 (WT) & 55\, 985.55000  & 7.0 \\
\swift/XRT & 00032033031 (WT) & 55\,991.09231 & 6.7 \\
\xmm & 0672282901&  56022.95692 & 26.9 \\
\swift/XRT & 00032033032 (WT) & 56\,031.141159 & 4.3 \\

\hline
\end{tabular}}
\begin{list}{}{}
\item[$^{a}$] Mid-point of the observations.
\end{list}
\end{table}

The broadband emission of these objects and their flaring activity are
believed to be connected to their high dipolar and/or toroidal
magnetic field: this is indeed supported by the measurement of surface
dipolar $B$ field usually of the order of $10^{14}$--$10^{15}$\,G
(inferred through the assumption that, as ordinary pulsars, they are
spun down via magnetic dipolar losses: $B \sim 3.2\times10^{19}(P
\dot{P})^{1/2}$\,G, where $P$ is the spin period in seconds, $\dot{P}$
its first derivative and we assumed a neutron star mass and radius of
$R\sim10^6$~cm and $M\sim1.4~M_{\odot}$, respectively). However, the
recent detection of an SGR showing all the typical emission properties
defining a magnetar \citep{vdh10,esposito10}, but with an inferred
dipolar surface magnetic field  $<7.5\times10^{12}$\,G \citep{rea10},
has put into question the need of a high dipolar magnetic
field (namely higher than the quantum electron critical field $B_{\rm
  Q}=m_ec^2/e\hbar\sim 4.4\times10^{13}$\,G ) for an object to show magnetar-like
activity.

On 2011 July 14, a new SGR-like burst and associated outburst were
discovered by the \swift\, Burst Alert Telescope (BAT), and
followed soon after by all X-ray satellites \citep{cummings11}.  The
fast slew of the \swift\, X-ray telescope (XRT) promptly detected a
new bright X-ray source at $\rm RA: 18^h22^m18\fs00$, $\rm Dec:
-16\degr 04' 26\farcs8$ (J2000; $1\farcs8$ error at a 90\% confidence
level; \citealt{pagani11}), with a spin period of $P\sim8.43$\,s
\citep{gogus11}. The lack of an optical/infrared counterpart
\citep{bandyopadhyay11,rmi11,deugarte11}, as well as the
characteristics of the bursts, the X-ray spin period and its spectral
properties \citep{eri11,erit11,rei11}, led to its identification as a
new magnetar candidate \citep{cummings11,halpern11}.

After its discovery, many attempts were made to measure the spin
period derivative of \src\ \citep{gsk11,kuiper11,livingstone11} in
order to estimate its surface dipolar field. We present here the timing and spectral results
of the first 9 months of X-ray monitoring (\S\ref{xrayreduction} and
\S\ref{results}) of the new magnetar candidate \src, a detection of
its quiescent counterpart in archival data (\S\ref{rst}), as well as
upper limits on its emission in the optical and radio bands
(\S\ref{gtc} and \S\ref{gbt}). A detailed study of the SGR-like
bursts, precise X-ray position and pulse profile modeling will be
reported elsewhere (Kouveliotou et al. in preparation). Using our
timing and spectral results, we model the source outburst decay, and secular evolution, which resulted in an estimate of the
its real age and crustal toroidal field (\S7).

\section{X-ray observations and data reduction}
\label{xrayreduction}

In this study, we used data obtained from different satellites (see Table~\ref{obs-log} for a summary).  Observations and data analysis are briefly described in the following.

\subsection{Swift data}

The X-Ray Telescope (XRT; \citealt{burrows05}) on-board \swift\ uses a front-illuminated CCD detector sensitive to photons between 0.2 and 10 keV. Two main readout modes are available: photon counting (PC) and windowed timing (WT). PC mode provides two dimensional imaging information and a 2.5073\,s time resolution; in WT mode only one-dimensional imaging is preserved, achieving a time resolution of 1.766 ms.  

The XRT data were uniformly processed with \textsc{xrtpipeline} (version 12, in the \textsc{heasoft} software package version 6.11), filtered and screened with standard criteria, correcting for effective area, dead columns, etc. The source counts were extracted within a 20-pixel radius (one XRT pixel corresponds to about $2\farcs36$). For the spectroscopy, we used the spectral redistribution matrices in \textsc{caldb} (20091130; matrices version v013 and v014 for the PC and WT data, respectively), while the ancillary response files were generated with \textsc{xrtmkarf}, and they account for different extraction regions, vignetting and point-spread function corrections.

\begin{figure*}
\hspace{-1cm}
\includegraphics[height=8cm]{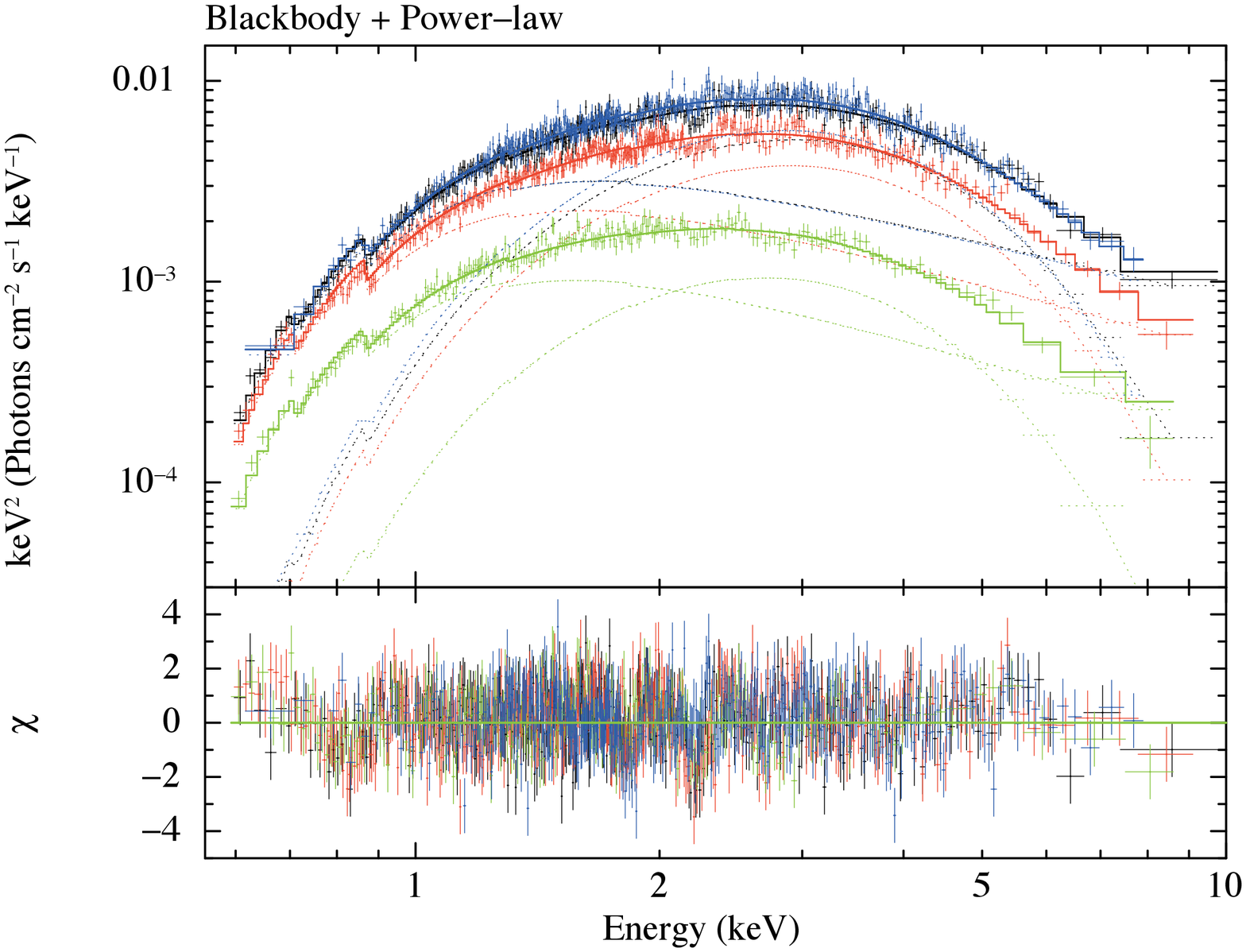}
\hspace{-1cm}
\includegraphics[height=8cm]{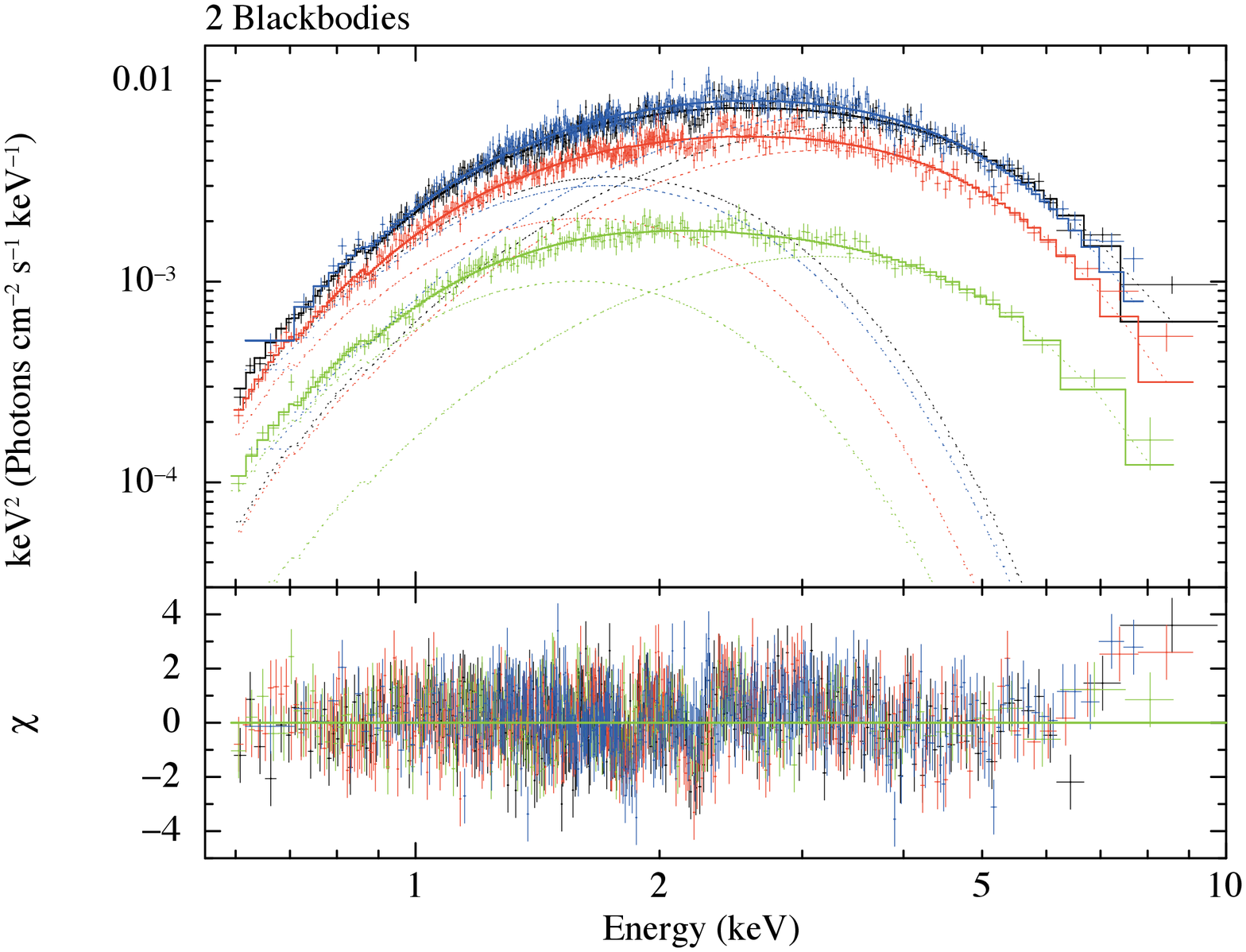}
\caption{Spectra of the \suzaku\, (blue) and \xmm\, observations fitted together with a blackbody plus power-law and two-blackbody models (see text and Table\,\ref{tab:bestspec} for details).  \smallskip}
\label{fig:bestspec}
\end{figure*}

\subsection{RXTE data}
The Proportional Counter Array (PCA; \citealt{jahoda96}) on-board \rxte\ consists of five collimated xenon/methane multianode Proportional Counter Units (PCUs) operating in the 2--60\,keV energy range.  Raw data were reduced using the \textsc{ftools} package (version 6.11). To study the timing properties of \src, we restricted our analysis to the data in Good Xenon mode, with a time resolution of 1 $\mu$s and 256 energy bins. The event-mode data were extracted in the 2--10\,keV energy range from all active detectors (in a given observation) and all layers, and binned into light curves of 0.1\,s resolution.

\begin{table*}
\caption{Spectral analysis of the \suzaku\, and \xmm\, data. }
\smallskip
\label{tab:bestspec}
\begin{center}
{\footnotesize
\begin{tabular}{@{}cccccccccc}
\hline
 & & & \multicolumn{3}{c}{Blackbody $+$ Powerlaw} &    \multicolumn{4}{c}{Two Blackbodies } \\
 Instrument & Time$^{a}$     &  Flux$^{b}$ &  kT~(keV)  &  R$_{\rm BB}$~(km)$^{c}$ &  $\Gamma$  & kT$_1$~(keV)    & $R_{\rm BB1}$~(km)$^{c}$  &   kT$_2$~(keV)    & $R_{\rm BB2}$~(km)$^{c}$   \\                   
\hline
\suzaku & 60.93$\pm$0.48  &   1.78$\pm$0.01  &  0.678$\pm$0.008 &  1.2$\pm$0.1 & 2.90$\pm$0.04 & 0.39$\pm$0.01 & 2.6$\pm$0.4 &  0.79$\pm$0.01 & 0.9$\pm$0.1  \\
\xmm & 70.25$\pm$0.06  &  1.70$\pm$0.01  &  0.689$\pm$0.006  &  1.1$\pm$0.1 & 2.86$\pm$0.03 & 0.40$\pm$0.01  & 2.6$\pm$0.4 &  0.84$\pm$0.01 & 0.8$\pm$0.1   \\
 \xmm & 90.06$\pm$0.11   &  1.20$\pm$0.03  &  0.679$\pm$0.005 &  1.0$\pm$0.1 & 2.99$\pm$0.03 & 0.37$\pm$0.02  & 2.5$\pm$0.4 & 0.79$\pm$0.01 & 0.7$\pm$0.1  \\
 \xmm & 266.1$\pm$0.2  &  0.4$\pm$0.1  &  0.623$\pm$0.008 &  0.61$\pm$0.08 & 3.05$\pm$0.04 & 0.35$\pm$0.01 & 2.0$\pm$0.2 & 0.78$\pm$0.01 &  0.4$\pm$0.1 \\
\hline
\end{tabular}}
\end{center}
\begin{list}{}{}
\item[$^{a}$] {\scriptsize Times are calculated in days from MJD 55\,757.0  .}
\item[$^{b}$] {\scriptsize  Fluxes are in units of $10^{-11}$\ergscm2 , referred to the BB+PL fit, and calculated in the 1--10\,keV energy range. Errors in the table are given at 1$\sigma$ confidence level. Reduced\,$\chi^2$ and absorption values are $\chi^2_\nu/$dof =1.05/2522 and $N_{\rm H}=0.50(1)\times10^{22}$\cm2 , $\chi^2_\nu/$dof =1.06/2522 and $N_{\rm H}=0.21(1)\times10^{22}$\cm2 , for the BB+PL and BB+BB models, respectively.}
\item[$^{c}$]  {\scriptsize Radii are calculated assuming a distance of 5\,kpc.}
\end{list}
\end{table*}


 \subsection{Suzaku data}

\suzaku\, \citep{mitsuda07} observed the field  of \src\, on 2012 September 13--14 with the pulsar located at the X-ray Imaging Spectrometer (XIS; \citealt{koyama07}) nominal position. The XIS consists of three front-illuminated (FI) CCD cameras (XIS0, XIS2 and XIS3), and one that is back-illuminated (BI; XIS1). One of the FI CCDs, XIS2, was not available at the time of our observation. XIS1 and XIS3 were operating in Normal Mode without any option (all the pixels on the CCD are read out every 8\,s), while XIS0 was operating with the 1/8 Window option allowing a read out time of 1\,s.

For each XIS, $3\times3$ and $5\times5$ edit modes cleaned event data were combined. Following standard practices, we excluded times within 436\,s of \emph{Suzaku} passing through the South Atlantic Anomaly and we also excluded the data when the line of sight was elevated above the Earth limb by less than $5^\circ$, or less than $20^\circ$ from the bright-Earth terminator. Moreover, we excluded time windows during which the spacecraft was passing through a cut-off rigidity of below 6\,GV. Finally, we removed hot and flickering pixels. The resulting total effective exposure was $\sim$33.5\,ks for each XIS. The SGR net count rates are 0.710(5), 1.180(6), and 1.060(6)\,count~s$^{-1}$ in the XIS0, XIS1 and XIS3, respectively. For the spectral analysis, we used only XIS0 and XIS3, which are the best calibrated cameras, while for the timing analysis we made use only of the XIS0 data, which owing to the 1/8 Window option, have a timing resolution adequate to sample the pulsar spin period.

\subsection{Chandra data}

The \chandra\, X-ray Observatory has observed \src\, with the High Resolution Imaging Camera (HRC--I; Zombeck et al. 1995) on 2011 July 28, for $\sim12$\,ks (ObsID: 13511).  Data were analyzed using standard cleaning procedures\footnote{http://asc.harvard.edu/ciao/threads/index.html} and {\tt CIAO} version 4.4. Photons were extracted from a circular region with a radius of 3$^{\prime\prime}$ around the source position, including more than 90\% of the source photons (see Kouveliotou et al. 2012 in prep for further details on this observation). We inferred an effective HRC--I count-rate of $70\pm1$\,counts\,s$^{-1}$.

\begin{figure*}
\centerline{
\includegraphics[height=7.5cm]{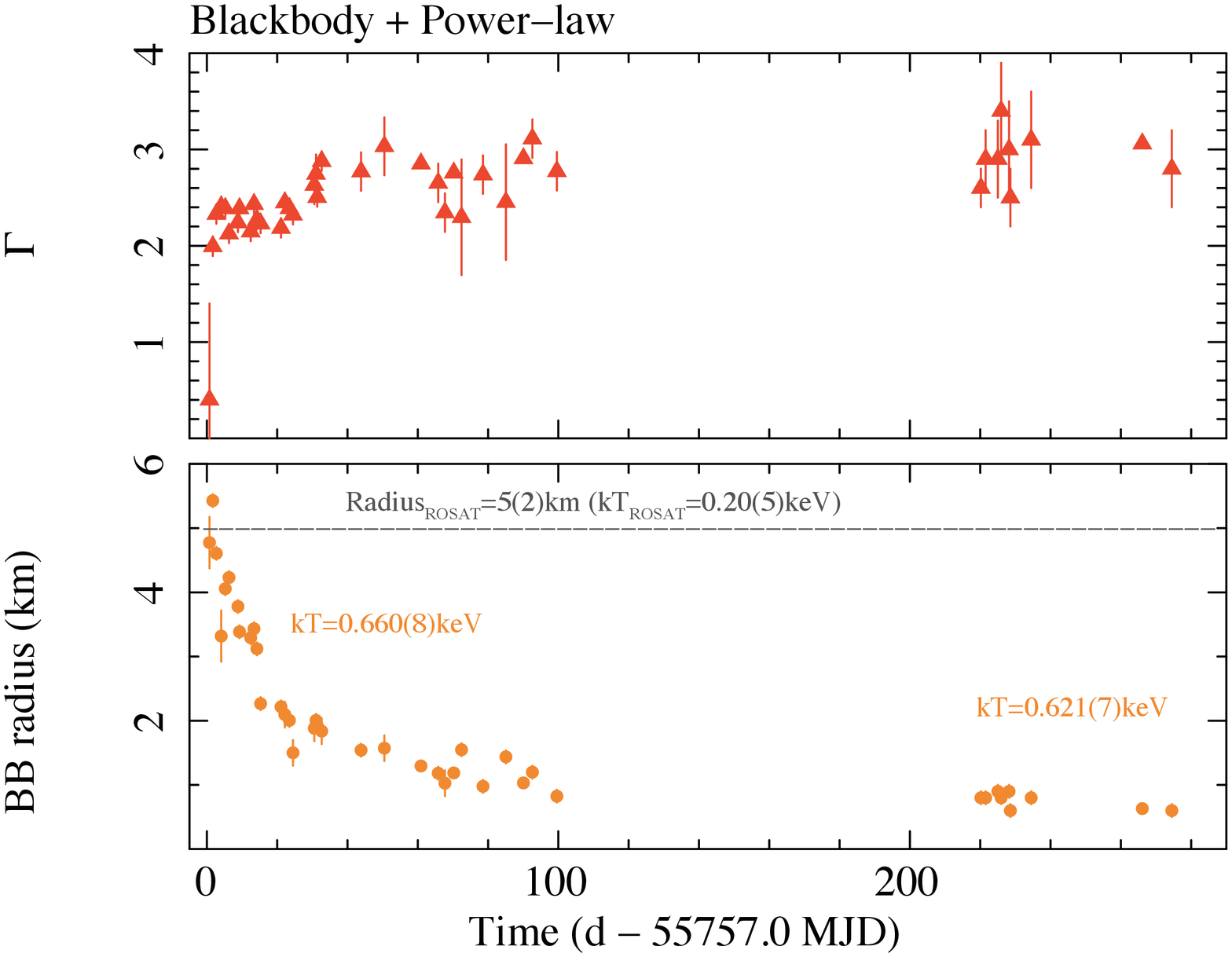}
\includegraphics[height=7.5cm]{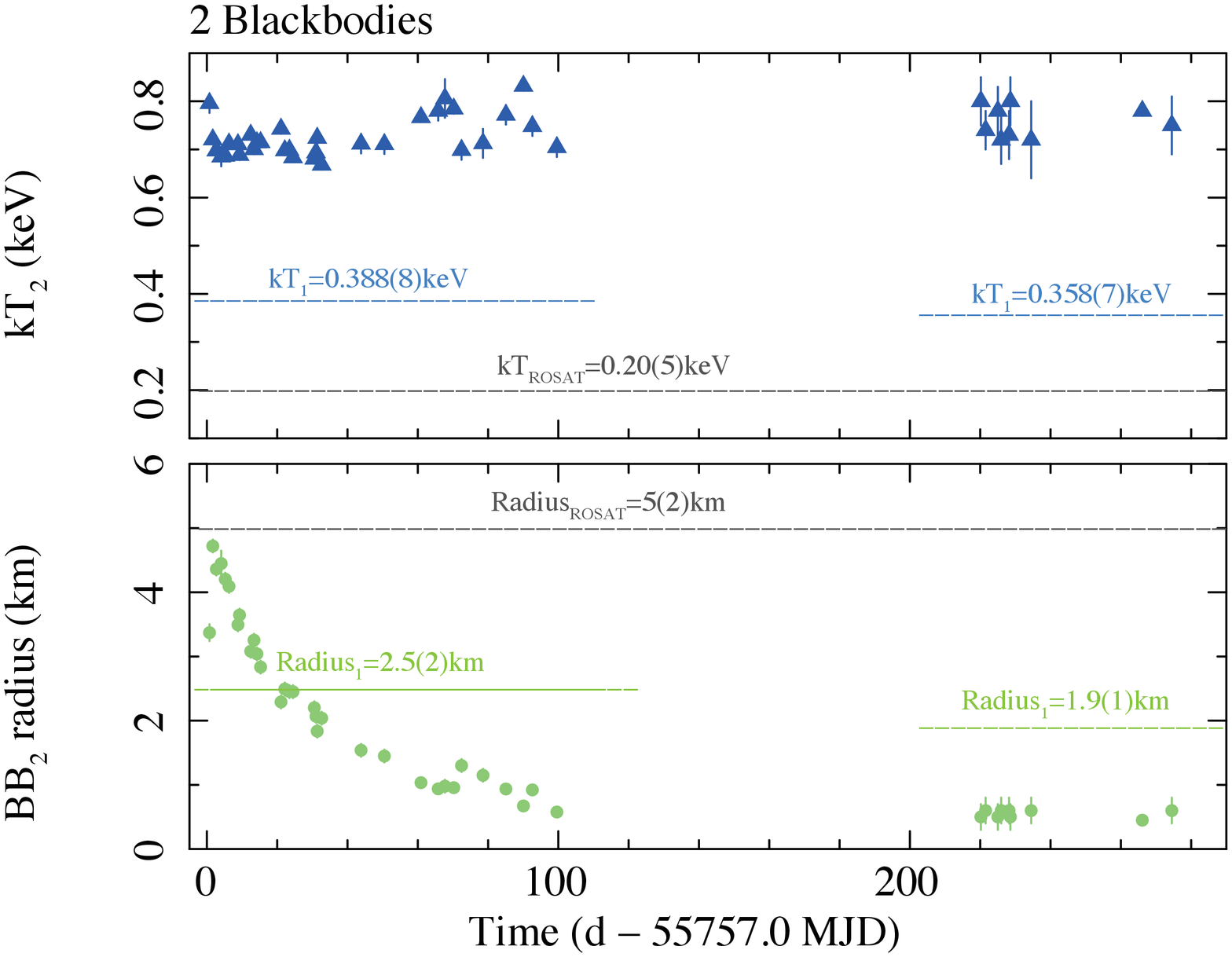}}
\caption{{\em Left panel}: Spectral parameters for a blackbody plus power-law fit.  {\em Right panel}:  Spectral parameters for a two-blackbody fit (see text and Table\,\ref{tab:spec} for details). }
\label{fig:spec}
\end{figure*}

\subsection{XMM-Newton data}
\label{xmm}

We observed \src\, three time with \xmm\ \citep{jansen01} on 2011 September 23, October 12, and April 05, for $\sim$10, 25, and 27 ks, respectively. Only the second observation was partially affected by background flares which we have cleaned during our spectral analysis (see \S\ref{spectral}) resulting in a net exposure time of 19.4\,ks. Data have been processed using SAS version 11, and we have employed the most updated calibration files available at the time the reduction was performed (April 2012). Standard data screening criteria are applied in the extraction of scientific products.  For our analysis we used only the EPIC-pn camera, and we checked that the two MOS cameras gave consistent results. The EPIC-pn camera was set in Prime Large Window mode (timing resolution 48\,ms), with the source at the aim-point of the camera. We have extracted the source photons from a circular region of 30\arcsec radius, and a similar region was chosen for the background in the same CCD but as far as possible from the source position. We restricted our analysis to photons having PATTERN$\leq$4 and FLAG=0. We find a EPIC-pn (background-subtracted) count rate of 5.03(3), 3.68(2) and 1.42(1)\,counts\,s$^{-1}$  for the first, second, and third observation, respectively.

\begin{table*}
\caption{Spectral analysis}
\smallskip
\label{tab:spec}
\begin{center}
{\footnotesize
\begin{tabular}{@{}ccccccc}
\hline
 & & & \multicolumn{2}{c}{Blackbody $+$ Powerlaw$^{*}$ } &    \multicolumn{2}{c}{Two Blackbodies$^{**}$ } \\
ObsID & Time$^{a}$     &  Flux$^{b}$ &   $\Gamma$      &   BB radius (km)$^{c}$  &   kT$_2$ (keV)    & BB$_2$ radius (km)$^{c}$   \\                   
\hline
33001 & 0.76$\pm$0.01  &  23.4$\pm$1.0  &  0.4$\pm$1.0  &  4.8$\pm$0.4  &  0.80$\pm$0.02  &  3.37$\pm$0.13 \\
33002 & 1.69$\pm$0.01  &  24.1$\pm$0.5  &  2.0$\pm$0.1  &  5.4$\pm$0.1  &  0.72$\pm$0.01  &  4.7$\pm$0.1 \\
33003 & 2.70$\pm$0.01  &  18.8$\pm$0.6  &  2.3$\pm$0.1  &  4.6$\pm$0.1  &  0.70$\pm$0.01  &  4.4$\pm$0.1 \\
33005 & 4.11$\pm$0.01  &  18.8$\pm$0.6  &  2.4$\pm$0.1  &  3.3$\pm$0.4  &  0.68$\pm$0.02  &  4.4$\pm$0.2 \\
33006 &  5.25$\pm$0.20  &  15.8$\pm$0.5  &  2.4$\pm$0.1  &  4.0$\pm$0.1  &  0.69$\pm$0.01  &  4.2$\pm$0.1 \\
33007 &  6.31$\pm$0.08  &  17.8$\pm$0.7  &  2.1$\pm$0.1  &  4.2$\pm$0.1  &  0.71$\pm$0.01  &  4.1$\pm$0.1 \\
33008 &  8.86$\pm$0.07  &  12.5$\pm$0.5  &  2.2$\pm$0.1  &  3.8$\pm$0.1  &  0.71$\pm$0.01  &  3.5$\pm$0.1 \\
33009 & 9.29$\pm$0.04  &  12.2$\pm$0.4  &  2.4$\pm$0.1  &  3.4$\pm$0.1  &  0.69$\pm$0.01  &  3.6$\pm$0.1 \\
33010 & 12.50$\pm$0.04  &  11.2$\pm$0.4  &  2.1$\pm$0.1  &  3.3$\pm$0.1  &  0.73$\pm$0.01  &  3.1$\pm$0.1 \\
33011 &  13.40$\pm$0.07  &  10.2$\pm$0.3  &  2.4$\pm$0.1  &  3.4$\pm$0.1  &  0.70$\pm$0.01  &  3.2$\pm$0.1 \\
33012 &  14.24$\pm$0.04  &  10.0$\pm$0.5  &  2.2$\pm$0.2  &  3.1$\pm$0.1  &  0.71$\pm$0.01  &  3.0$\pm$0.1 \\
33013 &  15.27$\pm$0.01  &  9.3$\pm$0.1  &  2.2$\pm$0.1  &  2.3$\pm$0.1  &  0.71$\pm$0.01  &  2.8$\pm$0.1 \\
51001 & 21.11$\pm$0.10  &  7.0$\pm$0.4  &  2.2$\pm$0.1  &  2.2$\pm$0.1  &  0.74$\pm$0.01  &  2.3$\pm$0.1 \\
51002 &  22.19$\pm$0.16  &  6.3$\pm$0.3  &  2.4$\pm$0.1  &  2.0$\pm$0.2  &  0.70$\pm$0.01  &  2.5$\pm$0.1 \\
51003 & 23.50$\pm$0.08  &  6.3$\pm$0.2  &  2.4$\pm$0.1  &  2.0$\pm$0.1  &  0.70$\pm$0.01  &  2.4$\pm$0.1 \\
51004 & 24.50$\pm$0.14  &  6.1$\pm$0.3  &  2.3$\pm$0.1  &  1.5$\pm$0.2  &  0.68$\pm$0.01  &  2.4$\pm$0.1 \\
51006 & 30.59$\pm$0.01  &  4.5$\pm$0.2  &  2.6$\pm$0.2  &  1.9$\pm$0.2  &  0.68$\pm$0.01  &  2.2$\pm$0.1 \\
51007 & 31.07$\pm$0.07  &  4.1$\pm$0.2  &  2.7$\pm$0.2  &  2.0$\pm$0.1  &  0.69$\pm$0.02  &  2.1$\pm$0.1 \\
51008 & 31.40$\pm$0.01  &  4.0$\pm$0.2  &  2.5$\pm$0.1  &  1.9$\pm$0.1  &  0.72$\pm$0.01  &  1.8$\pm$0.1 \\
51009 & 32.67$\pm$0.01  &  3.5$\pm$0.3  &  2.9$\pm$0.1  &  1.8$\pm$0.2  &  0.67$\pm$0.01  &  2.0$\pm$0.1\\
33015 &  43.87$\pm$0.10  &  2.7$\pm$0.3  &  2.8$\pm$0.2  &  1.5$\pm$0.1  &  0.71$\pm$0.02  &  1.5$\pm$0.1 \\
33016 & 50.49$\pm$0.31  &  2.4$\pm$0.4  &  3.0$\pm$0.3  &  1.6$\pm$0.1  &  0.71$\pm$0.02  &  1.4$\pm$0.1 \\
6002010 & 60.93$\pm$0.48  &  1.78$\pm$0.03  &  2.85$\pm$0.03  &  1.30$\pm$0.01  &  0.77$\pm$0.01  &  1.03$\pm$0.02 \\
33017 & 65.83$\pm$0.13  &  1.7$\pm$0.2  &  2.6$\pm$0.3  &  1.2$\pm$0.1  &  0.78$\pm$0.02  &  0.94$\pm$0.05 \\
33018 & 67.72$\pm$0.20  &  2.1$\pm$0.2  &  2.3$\pm$0.2  &  1.0$\pm$0.1  &  0.81$\pm$0.04  &  1.0$\pm$0.1 \\
81801 & 70.25$\pm$0.06  &  1.70$\pm$0.03  &  2.76$\pm$0.03  &  1.19$\pm$0.01  &  0.78$\pm$0.01  &  0.96$\pm$0.02 \\
33019 & 72.45$\pm$0.35  &  1.8$\pm$0.1  &  2.3$\pm$0.6  &  1.5$\pm$0.1  &  0.70$\pm$0.02  &  1.3$\pm$0.1 \\
33020 & 78.54$\pm$0.46  &  1.8$\pm$0.1  &  2.7$\pm$0.2  &  1.0$\pm$0.1  &  0.71$\pm$0.03 &  1.1$\pm$0.1 \\
33021 & 85.06$\pm$0.38  &  1.48$\pm$0.15  &  2.4$\pm$0.6  &  1.4$\pm$0.1 &  0.77$\pm$0.02  &  0.94$\pm$0.06 \\
82701 & 90.06$\pm$0.11  &  1.20$\pm$0.03  &  2.91$\pm$0.03  &  1.03$\pm$0.01  &  0.83$\pm$0.01  &  0.67$\pm$0.02 \\
33022 & 92.62$\pm$0.38  &  1.4$\pm$0.1  &  3.1$\pm$0.2  &  1.2$\pm$0.1  &  0.75$\pm$0.02  &  0.92$\pm$0.07 \\
33024 & 99.57$\pm$0.37  &  0.98$\pm$0.06  &  2.8$\pm$0.2  &  0.8$\pm$0.1  &  0.70$\pm$0.02  &  0.57$\pm$0.04 \\
\hline 
33025  & 220.2$\pm$0.8  &  0.55$\pm$0.08  &  2.6$\pm$0.2  &  0.8$\pm$0.1 &  0.80$\pm$0.05  &  0.5$\pm$0.2 \\
33026 & 221.5$\pm$0.4  &  0.54$\pm$0.06  &  2.9$\pm$0.3  &  0.8$\pm$0.1 &  0.74$\pm$0.04  &  0.6$\pm$0.2 \\
33027 & 225.0$\pm$0.1  &  0.47$\pm$0.08  &  2.9$\pm$0.4  &  0.9$\pm$0.1   &  0.78$\pm$0.05  &  0.5$\pm$0.2 \\
33028 & 226.0$\pm$0.1  &  0.47$\pm$0.08  &  3.4$\pm$0.5  &   0.8$\pm$0.1  &  0.72$\pm$0.05  &  0.6$\pm$0.2 \\
33029 & 228.2$\pm$0.1  &  0.46$\pm$0.08  &  3.0$\pm$0.5  &  0.9$\pm$0.1   &  0.73$\pm$0.05  &  0.6$\pm$0.2 \\
33030  & 228.5$\pm$0.2  &  0.54$\pm$0.08  &  2.5$\pm$0.3  &  0.6$\pm$0.1   &  0.80$\pm$0.05  &  0.5$\pm$0.2 \\

33031  & 234.5$\pm$0.4 & 0.50$\pm$0.08 & 3.1$\pm$0.5  &   0.8$\pm$0.1   &  0.72$\pm$0.08  &  0.6$\pm$0.2 \\

82901 &  266.1$\pm$0.2 & 0.41$\pm$0.02 & 3.06$\pm$0.04  &  0.63$\pm$0.02   &  0.78$\pm$0.01  &  0.45$\pm$0.08 \\

33032 & 274.5$\pm$0.4 & 0.5$\pm$0.1 & 2.8$\pm$0.4  &  0.6$\pm$0.1      &  0.75$\pm$0.06  &  0.6$\pm$0.2 \\

\hline
\end{tabular}}
\end{center}
\begin{list}{}{}
\item[$^{*}$]  {\scriptsize The absorption value and the blackbody temperature were fixed to be the same for all spectra of the first and second set: $N_{\rm H}=0.50(1)\times10^{22}$\cm2 , $kT =0.660(8)$\,keV and $kT =0.621(7)$\,keV, for the first and second set of spectra, respectively. Reduced\,$\chi^2_\nu/$dof = 1.1/6501.}
\item[$^{**}$] {\scriptsize The absorption value, blackbody temperature and radius were fixed to be the same for all spectra of the first and second set: $N_{\rm H}=0.21(1)\times10^{22}$\cm2 , $kT_1=0.388(8)$\,keV and BB$_1$ Radius = 2.5(1)\,km, and $kT_1=0.358(7)$\,keV and BB$_1$ Radius = 1.9(1)\,km, for the first and second set of spectra, respectively. Reduced\,$\chi^2_\nu/$dof = 1.1/6542 .}
\item[$^{a}$] {\scriptsize Times are calculated in days from MJD 55\,757.0  .}
\item[$^{b}$] {\scriptsize  Fluxes are in units of $10^{-11}$\ergscm2 , referred to the BB+PL fit, and calculated in the 1--10\,keV energy range. Errors in the table are given at 1$\sigma$ confidence level.}
\item[$^{c}$]  {\scriptsize Radii are calculated assuming a distance of 5\,kpc.}

\end{list}
\end{table*}


\section{Results of the X-ray monitoring}
\label{results}

\subsection{X-ray spectral modeling}
\label{spectral}

Spectra were extracted as explained in \S\ref{xrayreduction}, and
rebinned in order to have at least 20 counts per bin in the \swift\,
spectra, and 50 counts per bin in the \suzaku\, and \xmm\, spectra.
We started our spectral analysis by fitting our higher quality spectra, those from the three \xmm/pn, and \suzaku/XIS03 observations, with several models (using XSPEC version 12.7.0; see Figure\,\ref{fig:bestspec} and Table~\ref{tab:bestspec}). Best fits were found using a
blackbody plus power-law (BB+PL; $\chi^2_\nu/$dof =
1.05/2522) and a 2 blackbodies (2BBs; $\chi^2_\nu/$dof =
1.06/2522) model, all corrected for the photoelectric absorption ({\tt phabs} model with solar abundances assumed from Anders \& Grevesse (1989) and photoelectric cross-section from Balucinska-Church \& McCammon (1998)). The hydrogen column density along the
line of sight was fixed to the same value for all of the spectra for a
given model. We obtained N$_{\rm H}=0.50(1)$ and
$0.21(1)\times10^{22}$\cm2 (errors in the text are at 1$\sigma$ level
unless otherwise specified), for the BB+PL and 2BBs model,
respectively (see Table~\ref{tab:bestspec}). In Figure~\ref{fig:bestspec} we show the residuals of this spectral modeling, and note that, although statistically the fits can be considered equally good, the BB+BB model departs from the data at higher energies.

Already from this first analysis it is evident how the spectrum is changing in time, although very slowly. 

We then expanded our spectral modeling by
fitting simultaneously all the \swift/XRT, \suzaku/XIS03,
and \xmm/pn spectra. Again both models gave satisfactory
fits (see Table~\ref{tab:spec}). The hydrogen column density along the
line of sight was fixed to the same values found from the modeling of our previous analysis. Table~\ref{tab:spec} summarizes the obtained spectral parameters.

For the BB+PL model, we first allowed all parameters to vary freely,
and we noticed that the BB temperature was consistent with being
constant in time in the early phases of the outburst (most probably changing too little for our spectral analysis to be sensible to its variations). This was visible already  from Table~\ref{tab:bestspec} when considering only the most detailed spectra. We then tied the BB temperature across all spectra in the first 100 days of the outburst, and similarly we did for the last spectra (between 200-300\,days after the trigger). Best fit (reduced $\chi^2_\nu/$dof =
1.1/6501) was found with a BB temperature of $kT =
0.660(8)$\,keV and $kT =0.621(7)$\,keV, for the early and late times spectra, respectively. More detailed spectra would have certainly disentangled a slow decay between those two values. Figure~\ref{fig:spec} (left panel) shows the time
evolution of the power-law index ($\Gamma$) and the BB area. We can see how the latter shrinks as the
outburst decays, while the power law index increases slowly,
anti-correlated with the X-ray flux (see also Figure\,\ref{fig:phases}
left panel).

On the other hand, for the BB+BB model we noticed that, by leaving 
all the parameters free to vary across all the spectra, the
temperature and the radius of the first blackbody were not varying
significantly in time. Similarly to the BB+PL case, we then fixed those values to be the same in all
spectra at early and late times separately. This resulted in the best fit values of $kT_1=0.388(8)$ keV
and BB Radius$_1$ = 2.5(1)\,km, and $kT_1=0.358(7)$ keV
and BB Radius$_1$ = 1.9(1)\,km (to estimate the BB radii we assume a source distance of 5\,kpc). The best fit had a
reduced\,$\chi^2_\nu$/dof = 1.1/6542. The second blackbody has a
relatively steady temperature around $\sim$0.7 keV (see
Fig.~\ref{fig:spec} right panel) and its radius shrinks during the
outburst decay.

In the late time spectra, taken $>$200 days after the outburst
onset, the source flux decreases substantially (from $\sim24$ to
$0.4\times10^{-11}$\ergscm2 ), as the spectrum continues to soften.

We have also tried to model the spectra with a resonant cyclotron
scattering model \citep{rea07,rea08,zane09}, and, although the fits gave a
good chi-squared value ($\chi^2_\nu \sim$1.1/5912), the low magnetic
field of \src\, (see \S\ref{timing}) makes the use of those models,
envisaged for $B \sim 10^{14}$\,G, questionable (see also Turolla et
al. 2011).

We note that, although the $\chi^2$ values of both the BB+PL and BB+BB
fits might appear not acceptable from a purely statistical point of
view, many systematic errors are present in the simultaneous spectral
modeling of different satellites (the most severe being the
uncertainties in the inter-calibration between them, which is believed
to be within a 5\% error). We did not add any systematic error in the
spectral fitting to show the pure residuals of the fit; however, with
only 5\% systematic error, the reduced $\chi^2$ values would decrease
substantially, reaching a fully acceptable fit for both models
($\chi^2_\nu \sim$1.0).

\begin{figure*}
\vbox{
\hspace{-0.5cm}
\includegraphics[height=5cm]{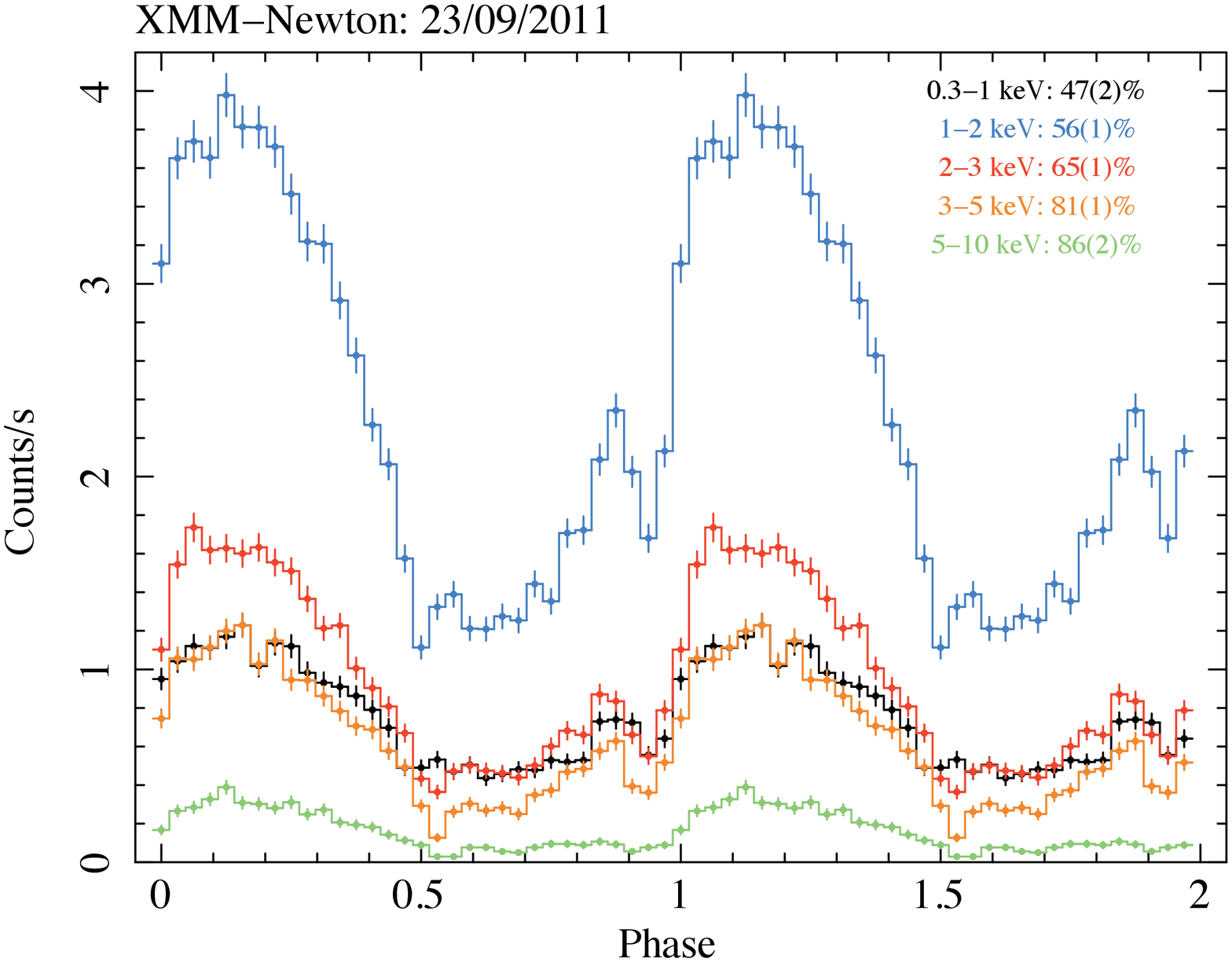}
\hspace{-0.5cm}
\includegraphics[height=5cm]{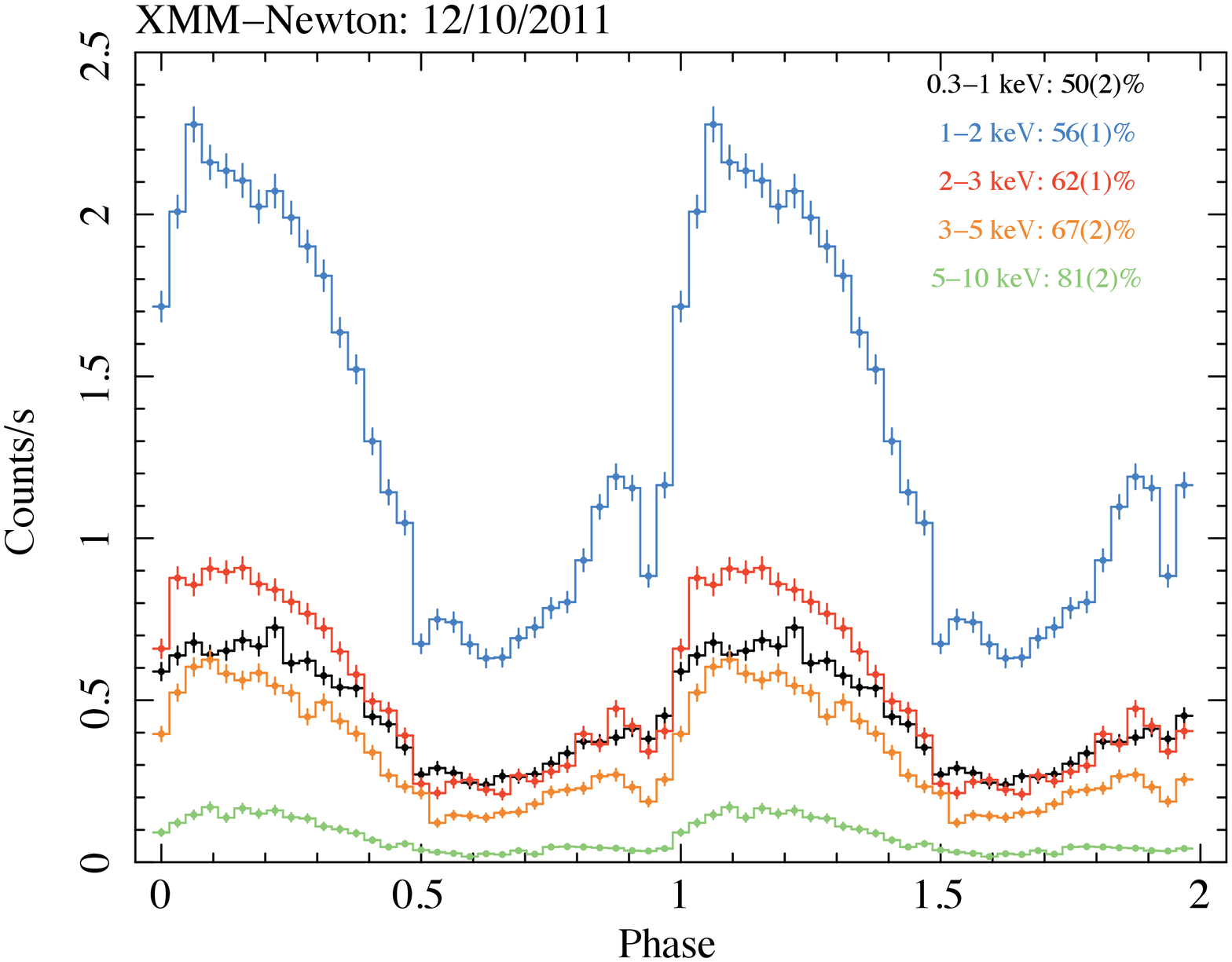} 
\hspace{-0.5cm}
\includegraphics[height=5cm]{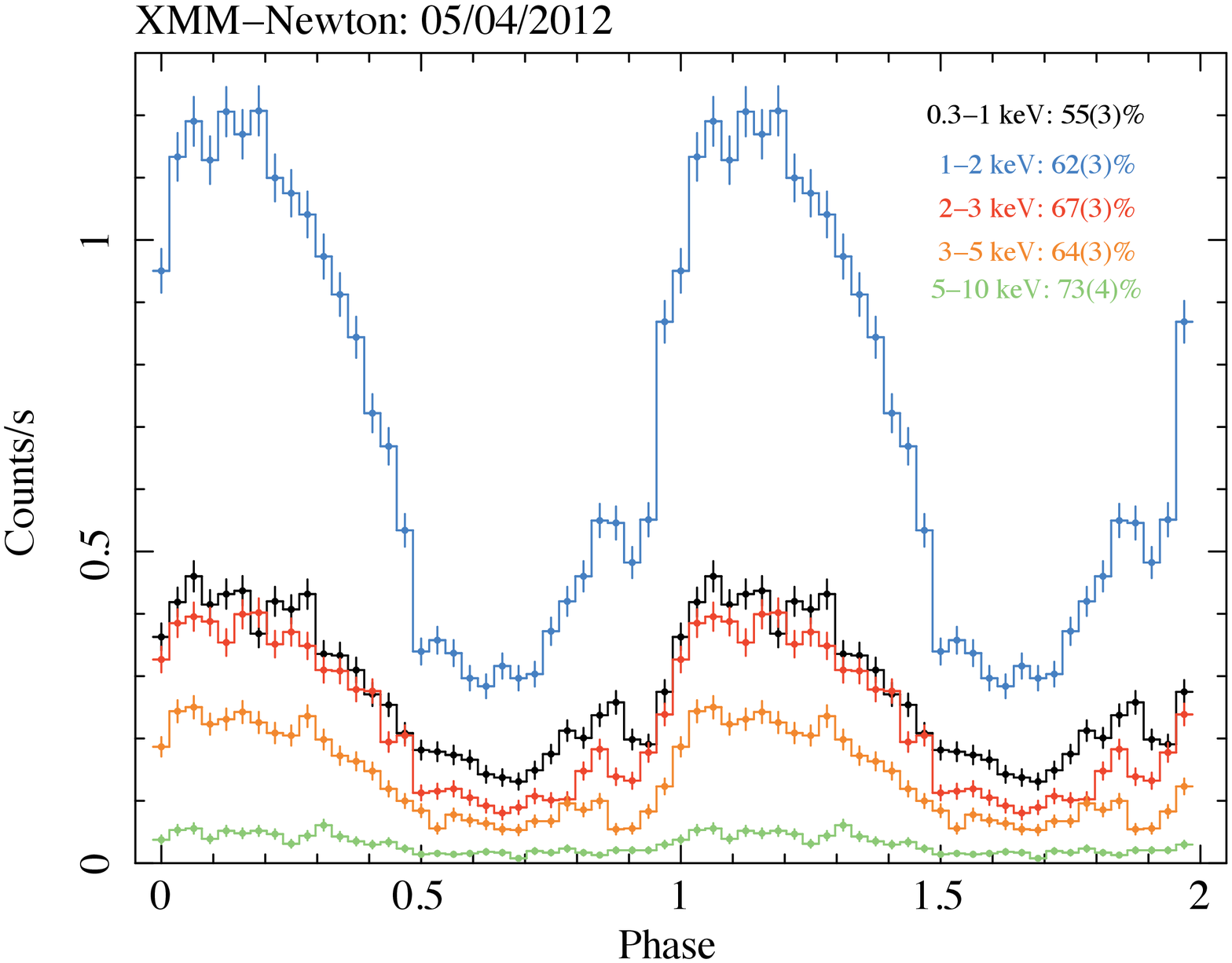} }
\caption{Pulse profiles with 32\,bins and relative pulsed fractions as a function of the energy band, for the three \xmm\, observations.}
\label{profiles}   
\end{figure*}

\subsection{X-ray timing analysis}
\label{timing}

For the X-ray timing analysis we used all data listed in Table\,\ref{obs-log}, after referring the event arrival times  to the barycenter of the Solar System (assuming the source coordinates by \citealt{pagani11} and the DE200 ephemeris for the Solar System).  The first \swift/XRT event lists were used in order to start building up a phase coherent timing solution and to infer the SGR timing properties. We started by obtaining an accurate period measurement by folding the data from the first two XRT pointings which were separated by less than 1\,day, and studying the phase evolution within these observations by means of a phase-fitting technique (see \citealt{dallosso03} for details). 
Due to the possible time variability of the pulse shape we decided not to use a pulse template in the cross-correlation, which might artificially  affect the phase shift, and we instead fit each individual folded light curve with two sine functions, the fundamental plus the first harmonic. 
In the following we also implicitly assume that the pulsation period  (and its derivative) is a reliable estimate of the spin period (and its derivative), an assumption which is usually considered correct for isolated neutron stars. 

The resulting best-fit period (reduced $\chi^2=1.1$ for 2 dof) is $P=8.43966(2)$\,s (all errors are given at 1$\sigma$ c.l.) at the epoch MJD 55757.0 . The above period accuracy of 20\,$\mu$s is enough to phase-connect coherently the later \swift, \rxte, \chandra, \suzaku, and \xmm\, pointings.  The procedure was repeated by adding, each time, a further observation folded at the above period, and following the phase evolution of the ascending node of the fundamental sine function best fitting the profile of each observation. The relative phases were such that the signal phase evolution could be followed unambiguously for the whole visibility window until November 2011 (see Figure\,\ref{fig:phases}). 
When adding the \rxte\, dataset, we also corrected the output phases by a small constant offset ($\sim$0.02), likely due to the different energy ranges and responses. 

We modeled the phase evolution with a polynomial function with a linear plus quadratic term, the inclusion of the latter results in a significant improvement of the fit (an F-test gives a probability of $7\times10^{-6}$ that the quadratic component inclusion is not required). The corresponding coherent solution (valid until November 2011) is  $P=8.43772007(9)$\,s and period derivative $\dot{P} = 1.1(2)\times 10^{-13}$\,s s$^{-1}$ ($\chi^2=132$ for 57 dof; at epoch MJD 55757.0). The above solution accuracy allows us to unambiguously extrapolate the phase evolution until the beginning of the next \swift\ visibility window which started in February 2012.

The final resulting phase-coherent solution (see also Table\,\ref{tab:timing}), once the latest 2012 observations are included, returns a best-fit period of $P=8.43772016(2)$\,s and period derivative of $\dot{P} = 8.3(2)\times 10^{-14}$\,s~s$^{-1}$  at MJD 55757.0 ($\chi^2=145$ for 67 dof; preliminary results were reported in \citealt{israel12}). The above best-fit values imply a surface dipolar magnetic field of $B\simeq2.7\times10^{13}$\,G (at the equator), a characteristic age of $\tau_{\rm c}=P/2\dot{P}\simeq1.6$\,Myr, and a spin-down power L$_{\rm rot}=4\pi I \dot{P}/P^3\simeq1.7\times 10^{30}$ \ergs (assuming a neutron star radius of 10\,km and a mass of 1.4$M_{\odot}$).

The final solution has a relatively high r.m.s. ($\sim$ 120\,ms) resulting in a best-fit reduced $\chi_{\nu}^2=2.1$. The introduction of high-order period derivatives in the fit of the phase evolution does not result in a significant improvement of the fit ($\chi^2=135$ for 66 dof; F-test gave a probability of $0.03$ that the cubic component inclusion is not required). This results in a 3$\sigma$ upper limit of the second derivative of the period of $\ddot{P}<5.8\times10^{-21}$s~s$^{-2}$ .

These values of $P$ and $\dot{P}$ are in agreement (within 1$\sigma$) with those inferred for the 2011 visibility window reported above\footnote{Note that the two data sets are not independent. However, we checked that when deriving a timing solution independently using only \rxte, \chandra, \xmm\, and \suzaku\, data for the first 120 days, and all the \swift\, observations plus the latest \xmm\, observation for the whole $\sim$300\, days, the two (now independent) solutions are still in agreement within 1$\sigma$.} However, this solution is not consistent, within $3\sigma$, with those already reported in the literature and based on a reduced dataset (\citealt{kuiper11,livingstone11}; valid until 70 and 90 days from the onset of the outburst, respectively). In order to cross check our results and to compare them with those previously reported we fit only those observations of our dataset within about 90 days from the trigger. The corresponding  best-fit parameters are  $P=8.4377199(1)$\,s and period derivative of $\dot{P} = 1.6(4)\times 10^{-13}$\,s s$^{-1}$ ($\chi^2=119$ for 52 dof; at epoch MJD 55757.0). The latter values are consistent with those of Livingstone et al. (2011). 

This analysis together with the relatively high r.m.s. value suggest that the timing parameters of the pulsar are  "noisy". Correspondingly, a timing solution based on a longer baseline may decrease the effect of a noisy behavior, while  those reported earlier are likely affected by the shorter time-scale variability of the timing parameters.    

In Figure\,\ref{profiles} we show the pulse profiles as a function of energy for the three \xmm\, observations (see \S\ref{xmm}), folded with the best-fit timing solution reported above. We derived pulsed fractions (defined as $\rm {PF = [Max(counts/s)-Min(counts/s)]/[Max(counts/s})+$ $\rm{Min(counts/s)]}$) in the 0.3--1, 1--2, 2--3, 3--5, and 5--10\,keV energy bands, of $47\pm2$, $56\pm1$, $65\pm1$, $81\pm1$, and $86\pm2$ \%, for the first observation; for the second observation: $50\pm2$, $56\pm1$, $62\pm1$, $67\pm2$, and $81\pm2$ \% ; and for the third observation: $55\pm3$, $62\pm3$, $67\pm3$, $64\pm3$, and $73\pm4$\% .

\section{\label{rst} ROSAT pre-outburst observations}

The \emph{R\"ontgensatellit} (\rosat) Position Sensitive Proportional Counter (PSPC; \citealt{pfeffermann87}) serendipitously observed the region of the sky including the position of \src\ between 1993 September 12 and 13 (obs. ID: rp500311n00), for an effective exposure time of 6.7 ks.

\begin{figure} 
\hspace{0.5cm}
\includegraphics[width=8cm]{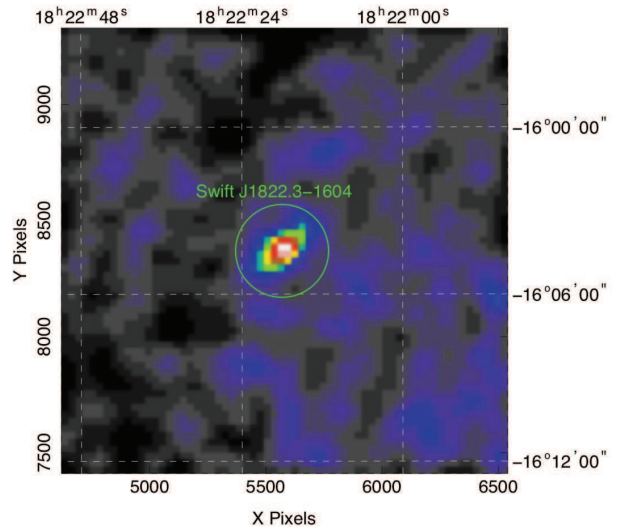}
\caption{\label{rosat} \rosat\ 1993 image of the region of \src. The circle is centred at the \swift/XRT position.}
\end{figure}

By means of a sliding cell source-detection algorithm, we found a 5.5$\sigma$ significant source with a 0.1--2.4 keV count rate of 0.012(3)counts s$^{-1}$ at the coordinates $\rm RA = 18^{h}22^{m}18\fs1,~Decl. = -16^\circ04'26\farcs4$ (positional uncertainty of 30\arcsec\ radius at 90\% c.l.; J2000). This source is also listed in the WGA and RXP catalogs, namely  1WGA\,J1822.2--1604 and 2RXP\,J182217.9--160417, with consistent values of  count rate. The positions of the latter objects are $\sim$20$''$ and 10$''$ from the \swift-XRT position. Given the relatively large \rosat/PSPC positional uncertainty, we believe the latter two sources and \src\, are the same object, which we propose as the SGR  quiescent counterpart (see Figure\,\ref{rosat}). 


We downloaded the relevant files of the ROSAT pointed observation and extracted the photon arrival times from a circle of $\sim100''$ radius (corresponding to an encircled energy of $>$90\%) around the X-ray position.  We found that an absorbed blackbody with \nh $< 7\times10^{21}$\,cm$^{-2}$ (see also \S\ref{spectral}), $kT = 0.20\pm0.05$ keV, and a radius of $5\pm2$\,km, best fit the data (reduced $\chi^2=0.8$ for 4 degree of freedom). We infer an observed flux of $\approx 1.3 \times10^{-13}$ \flux\ and  $\approx 4 \times10^{-14}$ \flux\ in the 0.1--2.4 keV  and 1--10\,keV energy ranges, respectively. Assuming a distance of 5\,kpc, this flux results in a bolometric luminosity during the quiescence state of $L_{\rm qui} \sim 4\times10^{32}$\ergs .

No significant periodic signal was found by means of a Fourier Transform, even restricting the search around the 8.44\,s period. The 3$\sigma$ upper limits on the pulsed fraction (semi-amplitude of the sinusoid divided by the source average count rate) is larger than 100\%.


\begin{table}
\caption{Timing parameters for \src.}
\begin{center}
{\small
\begin{tabular}{lr}
\hline
\hline
Reference Epoch (MJD) & 55757.0\\
Validity period (MJD) &  55757-- 56032\\
$P$\,(s) & 8.43772016(2) \\
$\dot{P}$\,(s\,s$^{-1}$) & $8.3(2)\times 10^{-14}$ \\
$\ddot{P}$\,(s\,s$^{-2}$) & $<5.8\times10^{-21}$ \\
$\nu$\,(Hz) & 0.118515426(3) \\
$\dot{\nu}$(s$^{-2}$)  & $1.17(3)\times10^{-15}$  \\
$\ddot{\nu}$(Hz\,s$^{-2}$)  & $<8.1\times10^{-23}$\\
$\chi^2/$dof & 145/66 \\
RMS residuals (ms) & 120 \\
\hline
B (Gauss) & $2.7\times10^{13}$ \\
L$_{\rm rot}$ (erg~s$^{-1}$) & $1.7\times
10^{30}$ \\
$\tau_{\rm c}$ (Myr) &  1.6 \\
\hline
\hline
\end{tabular}}
\end{center}
\label{tab:timing}
\end{table}


\begin{figure*}
\centerline{
\includegraphics[height=6cm]{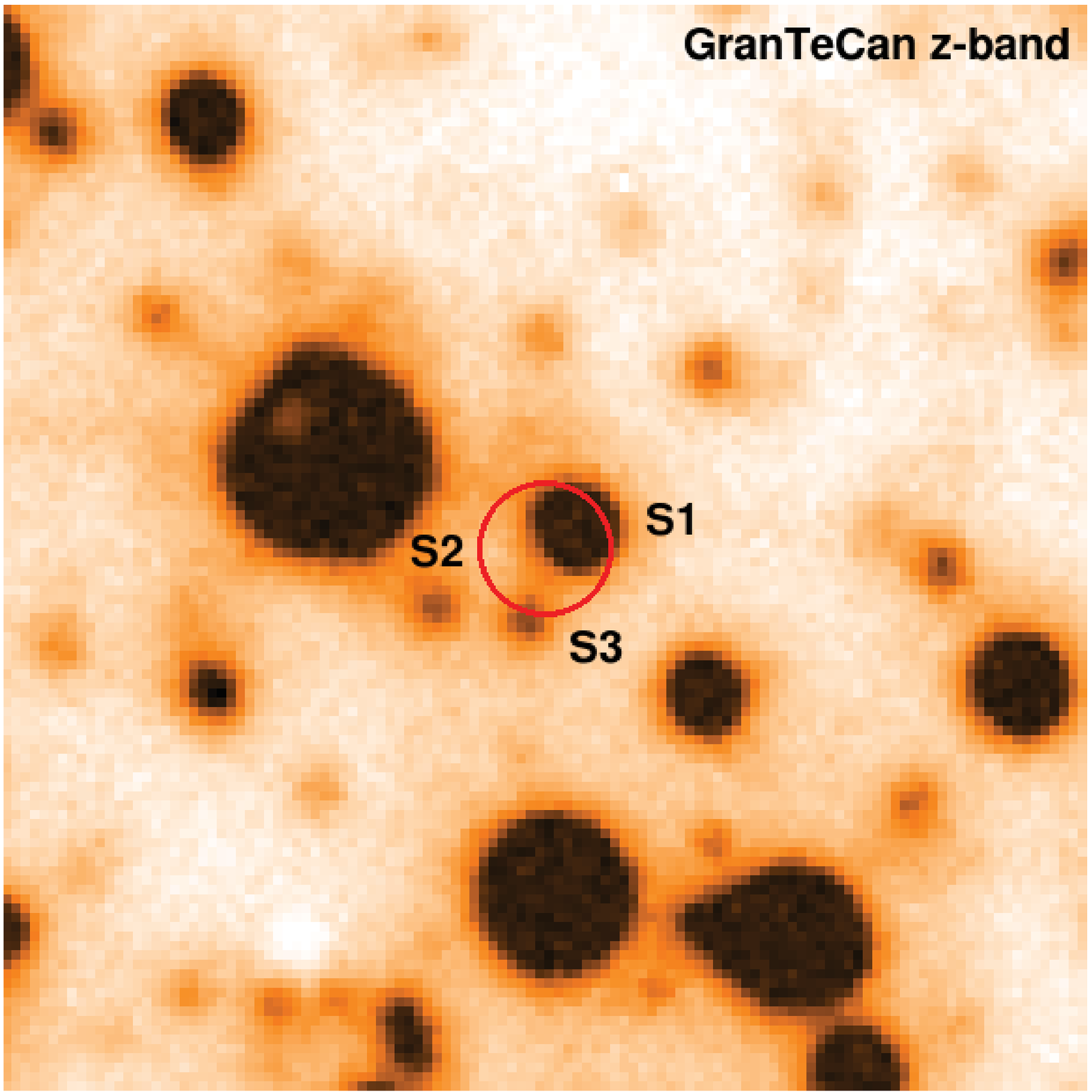} 
\hspace{1cm}
\includegraphics[height=6cm]{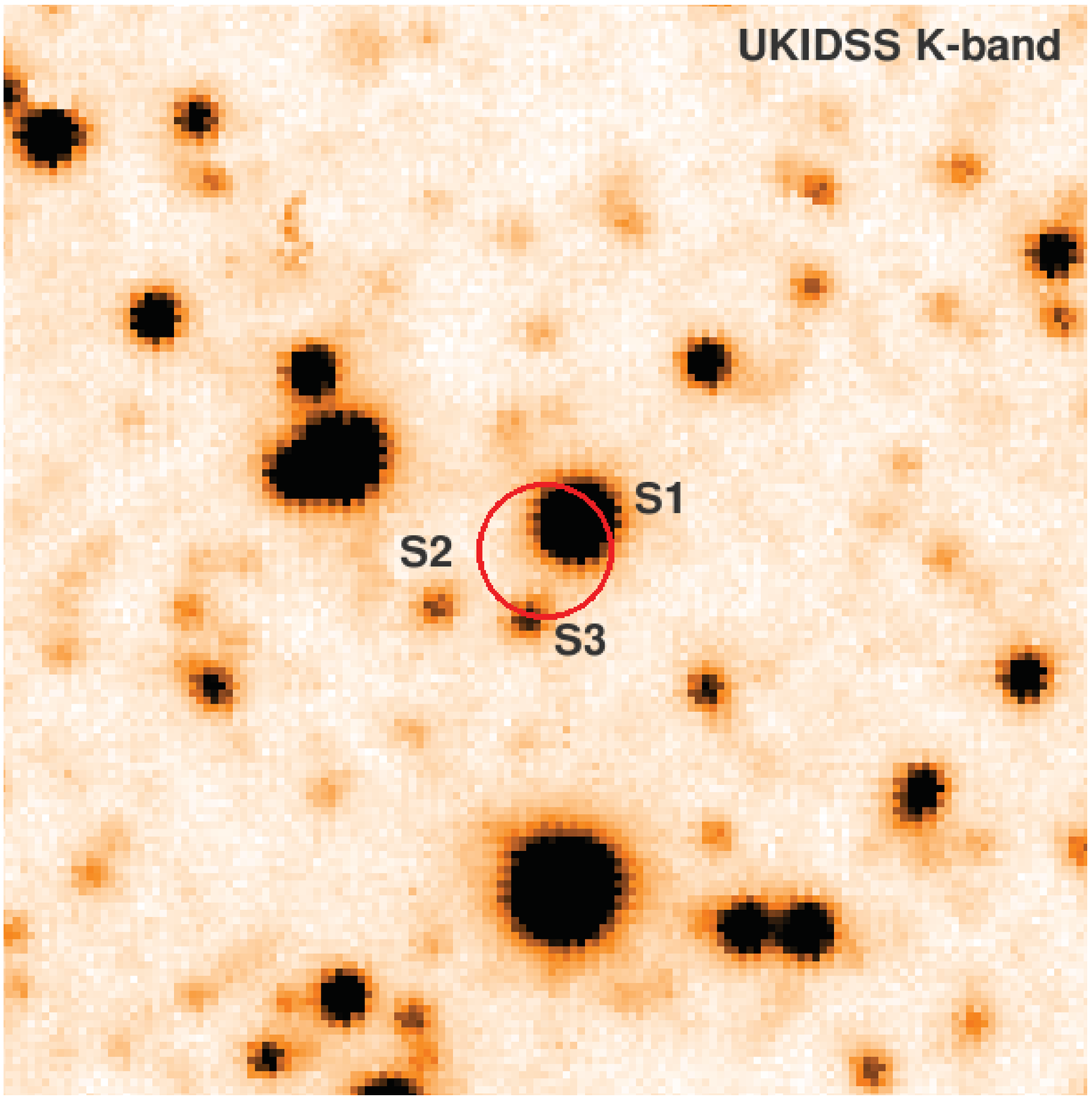} }
 \caption{{\em Left panel}: GranTeCan image of the \src\ field (30\arcsec $\times$ 30\arcsec) taken with the OSIRIS camera ($z$ band). The circle corresponds to the \swift/XRT position of the source \citep{pagani11}. The three objects detected close to, or within, the X-ray source error circle are labelled.  {\em Right panel:} UKIRT image of the same area obtained with the WFC in the $K$ band. }
\label{fc}      \end{figure*}

\section{Optical and infrared observations}
\label{gtc}

We observed the field of \src\ with the 10.4-m Gran Telescopio
Canarias (GranTeCan) at the Roque de los Muchachos Observatory (La
Palma, Spain). Images were taken in service mode on 2011 July 21 with
the OSIRIS camera, a two-chip Marconi CCD detector with
a nominal un-vignetted $7\farcm8\times7\farcm8$\arcmin\ field of view,
and an unbinned pixel size of $0\farcs125$. Observations were taken
through the Sloan $z$ filter ($\lambda=969.4$\, nm; $\Delta
\lambda=261$\, nm). We used a 5-point dithering pattern to correct for
the effects of the CCD fringing in the Red part of the spectrum. We
accurately selected the pointing of the telescope to position our
target in the right CCD chip and a bright ($B\sim 10$) star
$\sim$54\arcsec\ East of it in the left one, in order to avoid the
contamination from ghost images and saturation spikes. Unfortunately,
the observations were taken in conditions of very high sky background
due to the high lunar illumination, with the Moon phase at $\sim$0.5
and angular distance $\la 90^{\circ}$, and with a seeing ranging from
1--$2\farcs5$.  Observations were taken using exposure times of 108
and 54 s, with the latter chosen to minimize the sky background
induced by the Moon.  The total integration time was 4100 s. We
reduced the images with the dedicated tools in the IRAF {\tt ccdred}
package for bias subtraction and flat-fielding, using the provided
bias and sky flat images. We performed the photometric calibration
using exposures of the standard star PG\, 1528+0628.  In order to
achieve the highest signal--to--noise, we filtered out observations
taken with the highest seeing and sky background.  We aligned and
co-added all the best-images by means of the {\tt swarp} program
\citep{bertin02}, applying a $3 \sigma$ filter on the single pixel
average to filter out residual hot and cold pixels and cosmic ray
hits.  We performed the astrometry calibration of the OSIRIS image
with the {\tt WCStools} astrometry
package\footnote{http://tdc-www.harvard.edu/wcstools/}, using as a
reference the coordinates of stars selected from the GSC2 catalogue
\citep{lasker08}. Due to the significant and unmapped CCD distortions,
we only obtained an rms of $0\farcs3$ on the astrometric fit.  We
detected three objects (S1, S2, S3) within or close to the \src\,
position (see also \citealt{rmi11,gorosabel11}).  We computed their flux
through standard aperture photometry using the IRAF package {\tt
  apphot}. Their $z$-band magnitudes are $18.13\pm0.16$,
$20.05\pm0.04$, and $ 19.94\pm0.04$, for S1, S2 and S3, respectively
(see Figure\,\ref{gtc}). We detected no other object consistent with
the refined {\em Swift}/XRT position of \src\ \citep{pagani11} down to
a 3$\sigma$ limiting magnitude of $z=22.2\pm0.2$. Given their bright
optical magnitudes, we doubt that any of these objects is the optical
counterpart to \src. Based upon GranTecan spectroscopy, de Ugarte
Postigo \& Munos-Darias (2011) suggest that S1 and S2 are G to M-type
stars.

\label{ukidss}

As a reference, we inspected images of the \src\ field taken prior to our GranTeCan observations, i.e. when \src\ was probably in quiescence. To this aim, we searched for near-infrared (IR) observations taken as part of the UK  Infrared Deep Sky Survey (UKIDSS; \citealt{lawrence07}), performed with the Wide Field Camera (WFCAM;  \citealt{casali07}) at the  UK Infrared Telescope (UKIRT) at the Mauna Kea Observatory (Hawaii).   The \src\  field is indeed included in the UKIDSS Galactic Plane Survey (GPS) and data are available through Data Release 8 plus. Observations were taken on 2006 May 3rd  \citep{bandyopadhyay11}. We downloaded the fully reduced, calibrated, and co-added  $J$, $H$, $K$-band  science images of the \src\  field produced by the UKIDSS pipeline \citep{hambly08} together with the associated object catalogues through the WFCAM Science Archive (WSA)\footnote{http://surveys.roe.ac.uk/wsa/} interface.  The WFCAM astrometry is based on 2MASS \citep{skrutskie06} and is usually accurate to $\sim 0\farcs1$ \citep{lawrence07}.  We clearly identified objects S1 ($J=13.92$; $H=12.37$; $K=11.62$), S2 ($J=16.62$; $H=15.75$; $K=15.20$), S3 ($J=16.43$; $H=15.40$; $K=14.88$) in the UKIDSS images (see Figure\,\ref{ukidss}), with a relative flux comparable to the $z$-band flux measured on the OSIRIS ones.  No other object is detected at the \src\ position down to $5 \sigma$ limiting magnitudes of $J\sim 19.3$, $H\sim 18.3$ and $K\sim 17.3$.

\begin{figure*}
\centerline{
\includegraphics[width=18cm]{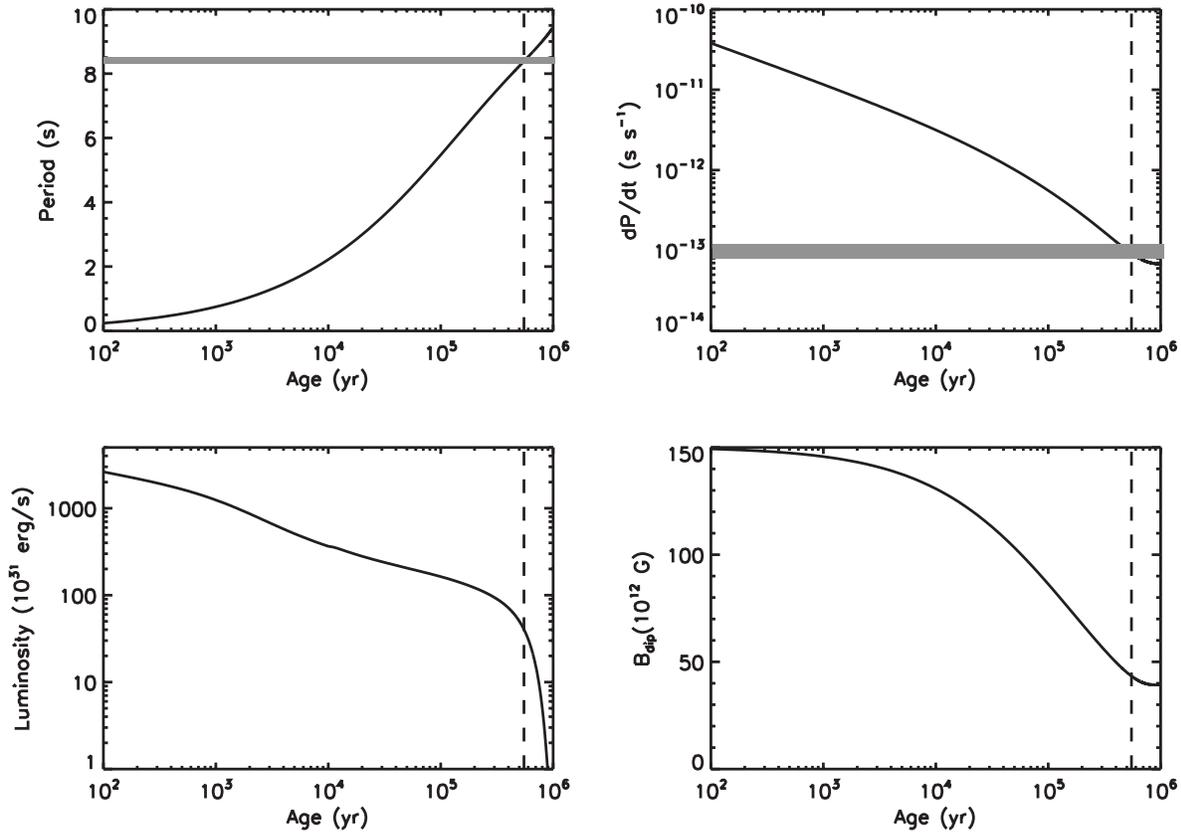}}
\caption{From top left to bottom right, the evolution of the period, period derivative, luminosity, and surface dipole field, according to the model discussed in the text. Grey lines report on the current values of the \src\, period and period derivative (with its error). The vertical dashed lines mark the source "real" age of 550\,kyr .}
\label{evolution}
\end{figure*}


\begin{table}
\caption{GBT radio observations of \src.}
\begin{center}
{\footnotesize
\begin{tabular}{cccc}
\hline
Date  & Start time (MJD)  & Exposure (s) & S$_{min}$~(mJy)$^{*}$  \\
\hline
2011-07-22 & 55764.26856481 & 1031.8269 & 0.06 \\
2011-08-18 & 55792.00894675 &  967.3408 & 0.06 \\
2011-09-20 & 55824.04596064 & 1365.0146 & 0.05 \\
2011-10-19 & 55853.87988425 & 1375.7644 & 0.05 \\
\hline
\end{tabular}}
\begin{list}{}{}
\item[$^{*}$]  S$_{min}$ is the minimum flux density reached.
\end{list}
\end{center}

\label{tab:radio}
\end{table}


\section{Radio observations}
\label{gbt}

Radio observations of \src\, were performed at the 101\,m Green Bank Telescope (GBT) on four occasions after the X-ray outburst, spaced by about a month one from the other (see Table \ref{tab:radio}). Data were acquired with the Green Bank Ultimate Pulsar Processing Instrument (GUPPI; \citealt{duplain08}) at a central frequency of 2.0\,GHz over a total observing bandwidth of 800\,MHz.

For each observation, in order to correct for the dispersive effects of the interstellar medium, the bandwidth was split into 1024 channels about 250 of which were unusable because of radio frequency interferences (RFI), leaving us with 600\,MHz of clean band. The observations lasted 16 to 23 minutes and were sampled every 0.6557\,ms. Since the pulsar rotational parameters are known from X-ray observations (see \S\ref{timing}), we first folded the data at the known period. We also folded the data at half, one third and a quarter of the nominal period in order to detect putative higher harmonics components of the intrinsic signal, in case the latter were deeply contaminated by RFI. Folding was done using \textsc{dspsr} \citep{vanstraten11} to form 30\,s long sub-integrations subdivided into 512 time bins. The sub-integrations and the 1024 frequency channels, cleaned from RFI, were then searched around the nominal period $P$ and over a wide range of dispersion measure (DM) values (from 0 to 1000\,pc\,cm$^{-3}$) to find the $P$--DM combination maximizing the signal-to-noise ratio. No dispersed signal was found in the data down to a flux of about 0.05\,mJy depending on the observation (see Table \ref{tab:radio}).

Data were also blindly searched using the code suites \textsc{presto}\footnote{See \url{http://www.cv.nrao.edu/$\sim$sransom/presto/}.} and \textsc{sigproc}\footnote{See http://sigproc.sourceforge.net/.}. In both cases, after de-dispersion of the data with 839 trial DMs (ranging from 0 to 1000\,pc\,cm$^{-3}$) and removal of the frequency channels affected by RFI, the time series are transformed using fast Fourier algorithms and their power spectra searched for relevant peaks. These Fourier-based search techniques require a $2^n$ number of time samples in input; for this reason the amount of data analyzed was 1030\,s, (a minute of fake data were added to the shortest observation) about two-thirds of the total of the longest observation, hence the flux limit attained, depending on the inverse of the square root of the integration time, was proportionally higher. With \textsc{sigproc} we also searched the data for single de-dispersed pulses but no signal was found in either the Fourier domain or the single pulse searches.

\section{Discussion}
\subsection{The secular thermal evolution of \src}
\label{sec:evolution}

To investigate whether the observed properties of \src\ are consistent with those
of an evolved magnetar, as suggested by its characteristic age of $\tau_{\rm c}\simeq1.6$\,Myr, we followed the secular evolution
of this object using a two-dimensional
magneto-thermal evolution code. We refer to Pons et al. (2009) and
Aguilera et al. (2008) for details about the code and the microphysical
inputs of the model.
This allows us to estimate the natal properties of the neutron star, its current age and internal field
strength.
We have considered the evolution, including magnetic field decay and heating by Ohmic diffusion, of an ultra-magnetized neutron star with a mass of 1.4~M$_{\odot}$, with no exotic phases nor fast neutrino cooling processes,
but with enhanced neutrino emission from the breaking and formation of neutron and
proton Cooper pairs (standard cooling scenario).
We assumed an initial neutron star spin
period of 10\,ms and an initial dipolar field of $B_{\rm dip} (t=0) =
1.5 \times 10^{14}$\,G.
In Figure \ref{evolution} we plot the evolution of spin period, period derivative, luminosity, and the dipolar
surface magnetic field of a model that can match the current observed
values at the ``real'' age of 550\, kyr.
The model has an initial crustal toroidal field that reaches a maximum value of $B_{\rm tor-max} (t=0) = 7\times10^{14}$\, G (approximately
half of the magnetic energy is stored in the toroidal component), which has now decayed to $B_{\rm tor-max}\sim1.3\times10^{14}$\, G. We have also studied the expected
outburst rate of this source, following the same procedure as in
\citet{perna11} and \citet{pons11}. We found that, at the present stage its outburst rate is very low
($\approx 10^{-3}$ yr$^{-1}$), because the magnetic field has been strongly dissipated.

\begin{figure*}
\centerline{
\includegraphics[width=10cm]{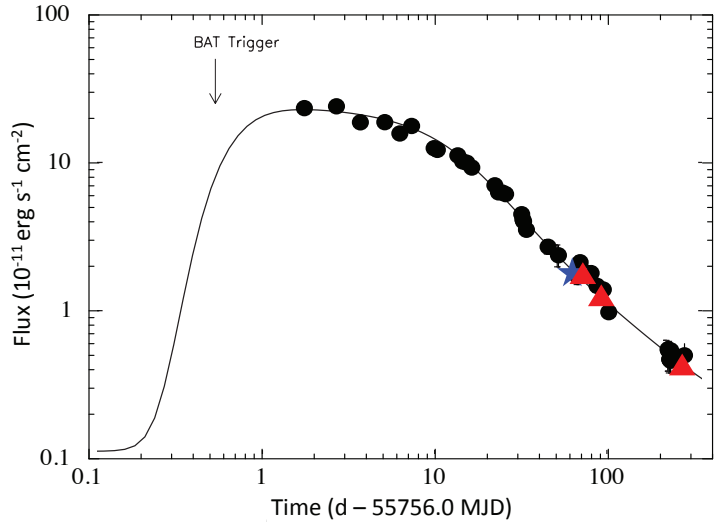}}
\caption{Outburst model from \citet{pons12} superimposed to the 1-10\,keV flux decay of \src\, (see text for details).}
\label{outmodel}
\end{figure*}

\subsection{The spectral evolution during the outburst}

The spectral evolution during the outburst decay in \src\ bears
resemblance to that observed in other magnetars in outburst, notably
\xte\,\citep{hg05,ber11}, \sgre\, \citep{rea09}, \wes\, \citep{albano10}, and the ``low field'' \sgrf\,
\citep{esposito10,rea10,turolla11}. The decrease in flux appears, in
fact, to be associated to a progressive spectral softening. Despite
present data do not allow for an unambiguous spectral characterization
over the entire outburst, evidence for a slow spectral softening is present in both the BB+BB and BB+PL models. In this respect we note that data are not consistent with a BB+PL fit in which the PL index is frozen to the
same value in all observations (fitting the values of $\Gamma$ in
Table~\ref{tab:spec} and Figure~\ref{fig:spec} with a constant function gives a
reduced $\chi^2 > 10$).

A BB+PL spectrum is observed at soft X-ray energies in most magnetar sources, and is interpreted in terms of resonant cyclotron up-scattering of thermal surface photons by magnetospheric electrons in a twisted magnetosphere (e.g. \citealt{tlk02,nobili08}). In this framework, the evolution of \src\ is compatible with seed photons originating in a relatively small surface region which is heated by the (magnetic) event which gave rise to the outburst, magnetic energy release deep in the crust (as in \citealt{lyubarsky02}) and/or ohmic dissipation of magnetospheric currents \citep{beloborodov09}. The heated region shrinks and cools progressively during the period covered by our observations (the equivalent BB radius decreased from $\sim5$~km to $\sim1$~km; in the following we always assume a 5\,kpc distance) as residual heat is radiated away and the non-thermal component shows a progressive softening as the magnetosphere untwists.
 
On the other hand, the spectral evolution of the source can be also
accommodated in the framework of a BB+BB spectral decomposition. In
this model, the thermal emission is usually associated with two
regions of different temperature and size which were heated during the
outburst. It is well possible that a single heated region is actually
produced, but with a meridional temperature gradient, which can be
schematized as e.g. a hotter cap surrounded by a warm ring, similarly to the
case of \xte\, \citep{pg08}. The
absence of a non-thermal tail is not in contrast with the twisted
magnetosphere model if the twist is small and/or it affected only a
limited bundle of closed field lines (see e.g. \citealt{esposito10}),
especially if the surface field is low, as in the present source.

The archival \rosat\, observation show that, in quiescence, the source has a blackbody spectrum with $kT\sim 0.2$\,keV and $R\sim 5$\,km. Although the radius is somehow small, it is not unreasonable to associate the \rosat\, BB to thermal emission from the entire star surface, given the large errors and the uncertain distance determination.

If the outburst produced a heated region, which for concreteness we
take to be a two-temperature cap, during the decay we witnessed a
gradual shrinking of the hotter region (from $\sim5$~km to
$<1$~km). The warm ring also shrunk and cooled down slowly during the first 300\,days after the outburst. 
Given the very slow spectral evolution of this component, we could obtain a good spectral modeling by fixing its temperature and radius to be constant during the first 100 \,days of the outburst,  and again (at a different value) in the last 200-300 days (see Figure\,\ref{fig:spec}). This should be most probably interpreted as a gradual cooling which could not be followed in detail by the current observations, rather than a temperature jump.x

\begin{figure*}
\centerline{
\includegraphics[width=10cm]{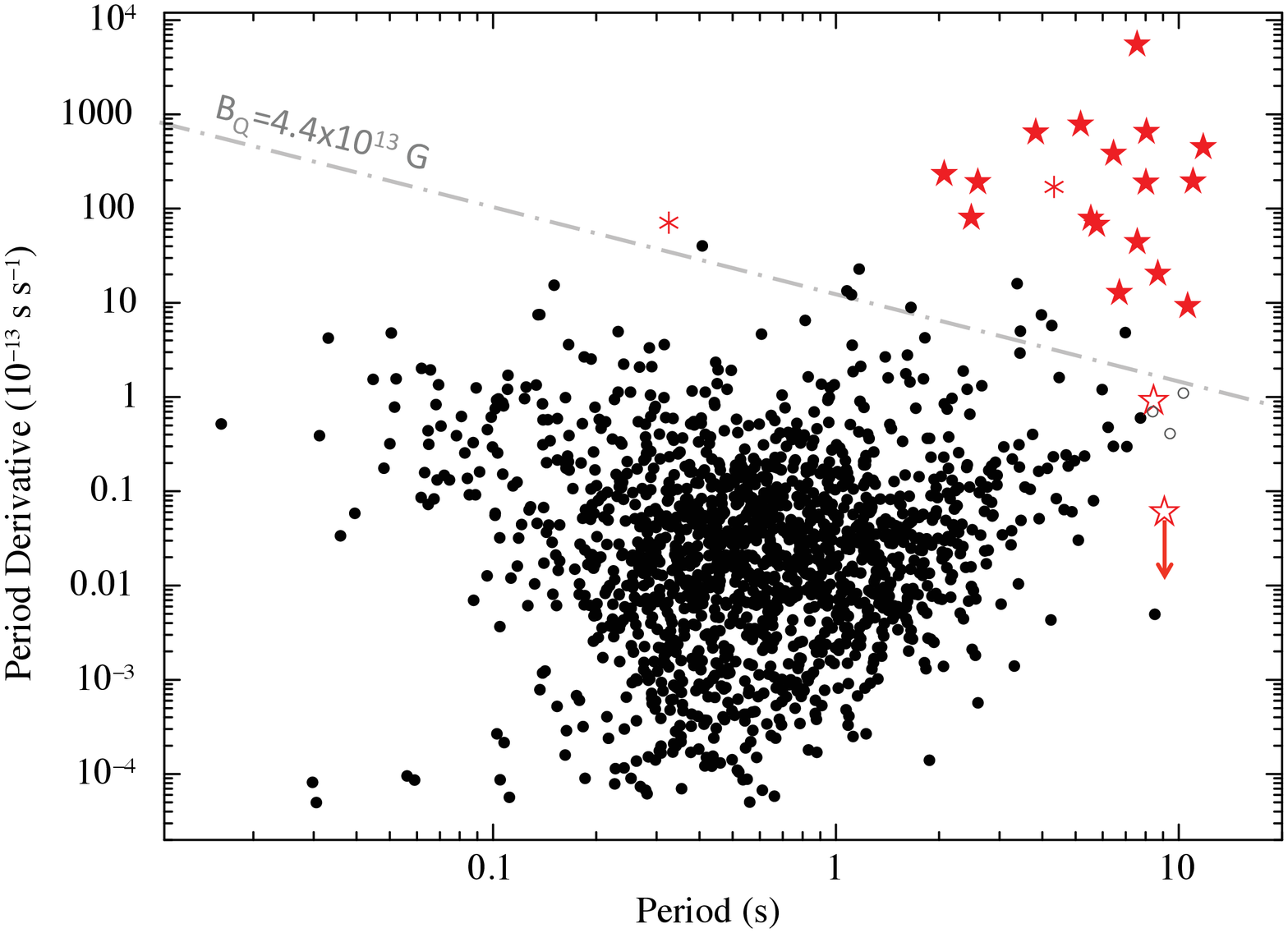}}
\caption{Period--period derivative diagram for all known isolated pulsars. Black dots are radio pulsars (from the ATNF Catalog; \citealt{manchester05}), while red symbols are all known magnetars. Asterics denote \hbpsr\, and \psr, and empty stars are \src\, and \sgrf. Empty grey circles are the X-ray Dim Isolated Neutron Stars (XDINS: \citealt{turolla09}). The dashed line represents the value of the critical electron magnetic field. }
\label{ppdot}
\end{figure*}

\subsection{The outburst decay and timescales}

The aggressive monitoring campaign we present here allowed us to study
in detail the flux decay of \src, and give an estimate of its typical
timescale.  Fitting the flux evolution in the first 225 days after the
onset of the bursting activity, we found that an exponential function
or a power law alone cannot fit the data properly, since at later time
(50--80 days) the decay slope starts to change. We found an acceptable
fit with an analytical function of the form $\mathrm{Flux}(t) = K_1 +
K_2e^{-(t/\tau)} + K_3t$ ($\chi^{2}$=4.7/37 dof); the best values of
the parameters are $K_1=(1.76\pm0.03)\times10^{-11}$\ergscm2 ,
$K_2=(22.0\pm0.3)\times10^{-11}$\ergscm2 , $\tau= 14.6\pm0.3$\,d, and
$K_3=(-5.2\pm0.2)\times10^{-14}$\ergscm2 d$^{-1}$. The outburst decays
of other magnetars are usually fitted by two components: an initial
exponential or power-law component accounting for the very fast
decrease in the first days or so (successfully observed only in very
few cases), followed by a much flatter power-law (see
\citealt{woods04,icd07,eiz08}). However, we note that the source has
not reached the quiescent level yet; hence the modeling of the outburst,
and relative timescale, might change slightly when adding further
observations until the complete quiescent level is reached.

We have also compared the observed outburst decay with the more physical theoretical
model presented in Pons \&  Rea (2012).  We have performed numerical
simulations with a 2D code designed to model the magneto-thermal evolution of neutron stars.
The pre-outburst parameters are fixed by fitting the timing properties to
the secular thermal evolution presented in section \S\ref{sec:evolution}.
We assume that \src\ is presently in an evolutionary state corresponding to that of the model presented in Figure\,\ref{evolution} at an age of 550\,kyr. We then model the outburst as the sudden release of energy in the crust, which is the progressively radiated away.
We have run several of such models varying the total injected energy (between
$10^{40}-10^{44}$\,erg), as well as the affected volume, which are the two
relevant parameters affecting the outburst decay
(coupled with the initial conditions which were explored in
\S\ref{sec:evolution}). The depth at which the energy is
injected and the injection rate bear less influence on the
late-time outburst evolution (Pons \&  Rea 2012).

In Figure\,\ref{outmodel} we show our best representative model that
reproduce the observed properties of the decay of \src\, outburst.  This model corresponds to an injection of $4\times10^{25}$~erg~cm$^{-3}$ in the outer crust, in the narrow layer
with density between $6\times10^{8}$ and $6\times 10^{10}$\,g~cm$^{-3}$, and in an angular region of 35 degrees (0.6 rad) around the pole. The total injected energy was then $1.3\times10^{42}$~erg . 

However, we must note that this solution is not unique and the parameter space is degenerate. Equally acceptable solutions
can be found varying the injection energy in the range $1-20\times10^{25}$~erg~cm$^{-3}$ and adjusting the other parameters.
The outer limit (low density) of the injection region affects the timescale of the rise of the light curve, which is probably too fast (1-10 hours) to be observable in most of the cases.
On the other hand, most of the light curve turns out to be insensitive to the inner limit (high density) of the injection
region. Only the outburst tail (at $>50$ days) is affected by this parameter, but this effect is hard to be distinguished from similar effects
from other microphysical inputs (e.g. varying the impurity content of the crust).
Finally, variations of the angular size can be partially compensated by
changes in the normalization factor which at present is undetermined (unknown
distance). This changes the volume implied and therefore
the estimate of the total energy injected.
Thus we need to wait for the full return to quiescence, and combine our
study with the complete analysis of the pulse profile and outburst spectrum, before we can place
better constraints on the affected volume and energetics.

\subsection{Radio and optical constraints}

A recent study on the emission of radio magnetars has shown that all
magnetars which exhibited radio pulsed emission, have a ratio of
quiescent X-ray luminosity to spin-down power L$_{\rm qui}/L_{\rm
  rot}<1$ \citep{rea12}. This suggests that the radio activity of
magnetars and of radio pulsars might be due to the same basic physical
mechanism, while its different observational properties are rather related
to the different topology of the external magnetic field (e.g. a
dipole and a twisted field; \cite{thompson08}.

In the case of \src , inferring the quiescent (bolometric) and spin-down
luminosities from our ROSAT data and our timing results (see
\S\ref{rst} and \S\ref{timing}), we derive L$_{\rm qui}/L_{\rm
  rot}\simeq4\times10^{32}$ \ergs $/ 1.7\times 10^{30}$ \ergs $\simeq 235$
. This value is in line with the source not showing any radio emission
(see \citealt{rea12} for further details).

Concerning the optical and infrared observations, the bright optical
fluxes of the sources S1--S3, much brighter than that of any other SGR
in outburst for a comparable distance and interstellar extinction, as well as
the lack of relative flux variability, suggest that objects S1--S3 are
most likely unrelated to \src.  The hydrogen column density derived
from the X-ray spectral fits ($N_{\rm H} = 0.5 \times
10^{22}$\,cm$^{-2}$) corresponds to an interstellar extinction
$E(B-V)\sim 0.89$ according to the relation of \citet{predehl95}.
Using the wavelength-dependent extinction coefficients of
\citet{fitzpatrick99}, this implies an absorption of $A_z \sim 1.33$
and $A_K \sim 0.32$ in the $z$ and $K$ band, respectively.  Our OSIRIS
upper limit would then correspond to an extinction-corrected spectral
flux of $\la 16.3 \mu$Jy in the $z$ band, or to an integrated flux
$\la 1.36 \times 10^{-14}$\,erg\,cm$^{-2}$\,s$^{-1}$.  For an assumed
\src\ distance of 5\,kpc, this implies an optical luminosity $L_z \la
4 \times 10^{31}$\,erg\,s$^{-1}$ during the outburst phase. Only very
few magnetars are detected in the optical band, the only ones being
\uu, \ee, and \sgre, all detected in the $I$ band (see,
e.g. \citealt{mignani11,dhillon11}).  The optical flux of, e.g. \ee\,
during its recent flaring phase \citep{wang08} corresponds to an
$I$-band luminosity $L_I \sim (4\pm 2) \times 10^{30}$\,erg\,s$^{-1}$
for an assumed distance of $3\pm 1$ kpc \citep{gmo05}.  Barring the
difference in comparing fluxes obtained through two slightly different
filters, our optical luminosity upper limit is about an order of
magnitude above the \ee\, luminosity.  Similarly, the flux upper limit
derived from the UKIDSS data implies for \src\ an IR luminosity $L_K
\la 1.1 \times 10^{32}$ \,erg\,s$^{-1}$ during the quiescent
phase. This upper limit is about an order of magnitude above the
computed luminosities of the magnetars' IR counterpart in quiescence
and, therefore, it is not very constraining.

\section{Conclusions}

We have reported on the outburst evolution of the new magnetar \src, which,
despite its relatively low magnetic field ($B=2.7\times10^{13}$\,G),
is in line with the outbursts observed for other magnetars with
higher dipolar magnetic fields (similar energetics, flux evolution and
spectral softening during the decay). Furthermore, we showed that the
non detection in the radio band is in line with its high X-ray
conversion efficiency ($L_{\rm qui}/L_{\rm rot}\simeq235$; see also
\citet{rea12} for further details).

We studied the secular thermal evolution of \src\, on the basis of the
actual value of its period, period derivative and quiescent
luminosity, and found that the current properties of the source can be reproduced if it has now an age of $\sim550$\,kyr, and it was born with a toroidal crustal field of $7\times10^{14}$\, G, which has by now decayed by less than an order of magnitude.

The position of \src\ in the $P$--$\dot P$ diagram (see
Figure\,\ref{ppdot}) is close to that of the ``low'' field magnetar
\sgrf\, \citep{rea10}. Although the fact that both have a sub-critical
dipole field is not relevant per se, and the dipolar field in \src\ is
at least four times higher than \sgrf, it is worth to stress that the
discovery of a second magnetar-like source with a magnetic field in
the radio-pulsar range strengthens the idea that magnetar-like
behavior may be much more widespread than what believed in the past,
and that it is related to the intensity and topology of the internal
and surface toroidal components, rather than only to the surface
dipolar field \citep{rea10,perna11,turolla11}.

Monitoring the source until its complete return to quiescence will be crucial to disentangle: 1) its complete spectral evolution during the outburst decay, 2) the possible presence of a second derivative of the rotational period, possibly due to the source timing noise, 3) refine further the modeling of the outburst and the surface region affected by this eruptive event.

\acknowledgements  We are indebted to the \swift, \rxte, \chandra, \suzaku\, and \xmm\, scheduling teams for the extraordinary job in promptly planning all the observations presented in this paper. We acknowledge the extraordinary support of the GTC staff, and in particular Rene Rutten and Carlos Alvarez for the prompt reaction to our ToO trigger. We also thank the GBT staff for scheduling these ToO observations so efficiently. NR is supported by a Ramon y Cajal Research Fellowship, and by grants AYA2009-07391, SGR2009-811, TW2010005 and iLINK 2011-0303. The Italian authors are supported by Agenzia Spaziale Italiana (ASI), Ministero dell'Istruzione, Universit\`a e Ricerca Scientifica e Tecnologica (MIUR -- COFIN), and Istituto Nazionale di Astrofisica (PRIN-INAF) grants. PE acknowledges financial support from the Autonomous Region of Sardinia through a research grant under the program PO Sardegna FSE 2007--2013, L.R. 7/2007 ``Promoting scientific research and innovation technology in Sardinia''. CK was partially supported by NASA grant NNH07ZDA001-GLAST.

%

\begin{thebibliography}{61}
\expandafter\ifx\csname natexlab\endcsname\relax\def\natexlab#1{#1}\fi

\bibitem[{{Aguilera} {et~al.}(2008){Aguilera}, {Pons}, \&
  {Miralles}}]{aguilera08}
{Aguilera}, D.~N., {Pons}, J.~A., \& {Miralles}, J.~A. 2008, \aap, 486, 255

\bibitem[{{Albano} {et~al.}(2010){Albano}, {Turolla}, {Israel}, {Zane},
  {Nobili}, \& {Stella}}]{albano10}
{Albano}, A., {Turolla}, R., {Israel}, G.~L., {Zane}, S., {Nobili}, L., \&
  {Stella}, L. 2010, \apj, 722, 788


\bibitem[{{Anders \& Grevesse}(1989){Anders} \& {Grevesse}}]{angr} 
{Anders}, E. \& {Grevesse}, N., 1989,  Geochimica et Cosmochimica Acta 53, 197 

\bibitem[{{Balucinska-Church \& McCammon}(1998){Balucinska-Church} \& {McCammon}}]{bcm98} 
{Balucinska-Church} \& {McCammon}, 1998, \apj, 496, 1044


\bibitem[{{Bandyopadhyay} {et~al.}(2011){Bandyopadhyay}, {Lucas}, \&
  {Maccarone}}]{bandyopadhyay11}
{Bandyopadhyay}, R.~M., {Lucas}, P.~W., \& {Maccarone}, T. 2011, Astron. Tel.,
  3502

\bibitem[{{Beloborodov}(2009)}]{beloborodov09}
{Beloborodov}, A.~M. 2009, \apj, 703, 1044


\bibitem[{{Bernardini} {et~al.}(2011){Bernardini}, {Perna}, {Gotthelf}, {Israel},
  {Rea}, \& {Stella}}]{ber11}
{Bernardini}, F., {Perna}, R., {Gotthelf}, E. G., {Israel}, G.~L., {Rea}, N.,
  \& {Stella}, L. 2011, \mnras, 418, 638


\bibitem[{{Bertin} {et~al.}(2002){Bertin}, {Mellier}, {Radovich}, {Missonnier},
  {Didelon}, \& {Morin}}]{bertin02}
{Bertin}, E., {Mellier}, Y., {Radovich}, M., {Missonnier}, G., {Didelon}, P.,
  \& {Morin}, B. 2002, in Astronomical Society of the Pacific Conference
  Series, Vol. 281, Astronomical Data Analysis Software and Systems XI, ed.
  {D.~A.~Bohlender, D.~Durand, \& T.~H.~Handley}, 228

\bibitem[{{Burrows} {et~al.}(2005){Burrows}, {Hill}, {Nousek}, {Kennea},
  {Wells}, {Osborne}, {Abbey}, {Beardmore}, {Mukerjee}, {Short}, {Chincarini},
  {Campana}, {Citterio}, {Moretti}, {Pagani}, {Tagliaferri}, {Giommi},
  {Capalbi}, {Tamburelli}, {Angelini}, {Cusumano}, {Br{\"a}uninger}, {Burkert},
  \& {Hartner}}]{burrows05}
{Burrows}, D.~N., et~al. 2005, Space Science Reviews, 120, 165

\bibitem[{{Casali} {et~al.}(2007){Casali}, {Adamson}, {Alves de Oliveira},
  {Almaini}, {Burch}, {Chuter}, {Elliot}, {Folger}, {Foucaud}, {Hambly},
  {Hastie}, {Henry}, {Hirst}, {Irwin}, {Ives}, {Lawrence}, {Laidlaw}, {Lee},
  {Lewis}, {Lunney}, {McLay}, {Montgomery}, {Pickup}, {Read}, {Rees}, {Robson},
  {Sekiguchi}, {Vick}, {Warren}, \& {Woodward}}]{casali07}
{Casali}, M., et~al. 2007, \aap, 467,
  777

\bibitem[{{Cummings} {et~al.}(2011){Cummings}, {Burrows}, {Campana}, {Kennea},
  {Krimm}, {Palmer}, {Sakamoto}, \& {Zane}}]{cummings11}
{Cummings}, J.~R., {Burrows}, D., {Campana}, S., {Kennea}, J.~A., {Krimm},
  H.~A., {Palmer}, D.~M., {Sakamoto}, T., \& {Zane}, S. 2011, Astron. Tel.,
  3488

\bibitem[{{Dall'Osso} {et~al.}(2003){Dall'Osso}, {Israel}, {Stella}, {Possenti}, {Perozzi}}]{dallosso03}
{Dall'Osso}, S., {Israel}, G. L., {Stella}, L., {Possenti}, A., {Perozzi}, E. 2003, \apj, 599, 485

\bibitem[{{de Ugarte Postigo} \& {Munoz-Darias}(2011)}]{deugarte11}
{de Ugarte Postigo}, A. \& {Munoz-Darias}, T. 2011, Astron. Tel., 3518

\bibitem[{{Dhillon} {et~al.}(2011){Dhillon}, {Marsh}, {Littlefair},
  {Copperwheat}, {Hickman}, {Kerry}, {Levan}, {Rea}, {Savoury}, {Tanvir},
  {Turolla}, \& {Wiersema}}]{dhillon11}
{Dhillon}, V.~S.,  et~al. 2011, \mnras,
  416, L16

\bibitem[{{DuPlain} {et~al.}(2008){DuPlain}, {Ransom}, {Demorest}, {Brandt},
  {Ford}, \& {Shelton}}]{duplain08}
{DuPlain}, R., {Ransom}, S., {Demorest}, P., {Brandt}, P., {Ford}, J., \&
  {Shelton}, A.~L. 2008, in Society of Photo-Optical Instrumentation Engineers
  (SPIE) Conference Series, Vol. 7019, Advanced Software and Control for
  Astronomy II, eds. {Bridger}, A. and {Radzwill}, N. Proceedings of the SPIE.
  SPIE, Bellingham WA, 70191D1--70191D10

\bibitem[{{Esposito} {et~al.}(2010){Esposito}, {Israel}, {Turolla}, {Tiengo},
  {G{\"o}tz}, {de Luca}, {Mignani}, {Zane}, {Rea}, {Testa}, {Caraveo}, {Chaty},
  {Mattana}, {Mereghetti}, {Pellizzoni}, \& {Romano}}]{esposito10}
{Esposito}, P., et~al. 2010, \mnras, 405, 1787

\bibitem[{{Esposito} {et~al.}(2008){Esposito}, {Israel}, {Zane}, {Senziani},
  {Starling}, {Rea}, {Palmer}, {Gehrels}, {Tiengo}, {De Luca}, {G{\"o}tz},
  {Mereghetti}, {Romano}, {Sakamoto}, {Barthelmy}, {Stella}, {Turolla},
  {Feroci}, \& {Mangano}}]{eiz08}
{Esposito}, P., et~al. 2008, \mnras, 390, L34

\bibitem[{{Esposito} {et~al.}(2011{\natexlab{a}}){Esposito}, {Rea}, \&
  {Israel}}]{eri11}
{Esposito}, P., {Rea}, N., \& {Israel}, G.~L. 2011{\natexlab{a}}, Astron. Tel.,
  3490

\bibitem[{{Esposito} {et~al.}(2011{\natexlab{b}}){Esposito}, {Rea}, {Israel},
  \& {Tiengo}}]{erit11}
{Esposito}, P., {Rea}, N., {Israel}, G.~L., \& {Tiengo}, A. 2011{\natexlab{b}},
  Astron. Tel., 3495

\bibitem[{{Fitzpatrick}(1999)}]{fitzpatrick99}
{Fitzpatrick}, E.~L. 1999, \pasp, 111, 63

\bibitem[{{Gaensler} {et~al.}(2005){Gaensler}, {McClure-Griffiths}, {Oey},
  {Haverkorn}, {Dickey}, \& {Green}}]{gmo05}
{Gaensler}, B.~M., {McClure-Griffiths}, N.~M., {Oey}, M.~S., {Haverkorn}, M.,
  {Dickey}, J.~M., \& {Green}, A.~J. 2005, \apjl, 620, L95

\bibitem[{{Gavriil} {et~al.}(2002){Gavriil}, {Kaspi}, \& {Woods}}]{gavriil02}
{Gavriil}, F.~P., {Kaspi}, V.~M., \& {Woods}, P.~M. 2002, \nat, 419, 142

\bibitem[{{Gogus} {et~al.}(2011){Gogus}, {Strohmayer}, \&
  {Kouveliotou}}]{gsk11}
{Gogus}, E., {Strohmayer}, T., \& {Kouveliotou}, C. 2011, Astron. Tel., 3503

\bibitem[{{Gorosabel} {et~al.}(2011){Gorosabel}, {de Ugarte Postigo},
  {Mottola}, {Hellmich}, {Ferrero}, {Sanchez-Ramirez}, {Tello}, {Casanova},
  {Marti}, \& {Castro-Tirado}}]{gorosabel11}
{Gorosabel}, J., et~al. 2011, Astron. Tel., 3496

\bibitem[{{G{\"o}{\u g}{\"u}{\c s}} \& {Kouveliotou}(2011)}]{gogus11}
{G{\"o}{\u g}{\"u}{\c s}}, E. \& {Kouveliotou}, C. 2011, Astron. Tel., 3542

\bibitem[{{Halpern} \& {Gotthelf} (2005)}]{hg05}
{Halpern}, J. P \& {Gotthelf}, E. V., 2005, \apj, 618, 874


\bibitem[{{Halpern}(2011)}]{halpern11}
{Halpern}, J.~P. 2011, GCN Circ., 12170

\bibitem[{{Hambly} {et~al.}(2008){Hambly}, {Collins}, {Cross}, {Mann}, {Read},
  {Sutorius}, {Bond}, {Bryant}, {Emerson}, {Lawrence}, {Rimoldini}, {Stewart},
  {Williams}, {Adamson}, {Hirst}, {Dye}, \& {Warren}}]{hambly08}
{Hambly}, N.~C., et~al. 2008, \mnras, 384,
  637

\bibitem[{{Ibrahim} {et~al.}(2004){Ibrahim}, {Markwardt}, {Swank}, {Ransom},
  {Roberts}, {Kaspi}, {Woods}, {Safi-Harb}, {Balman}, {Parke}, {Kouveliotou},
  {Hurley}, \& {Cline}}]{ibrahim04}
{Ibrahim}, A.~I., et~al. 2004, \apjl, 609, L21

\bibitem[{{Israel} {et~al.}(2007){Israel}, {Campana}, {Dall'Osso}, {Muno},
  {Cummings}, {Perna}, \& {Stella}}]{icd07}
{Israel}, G.~L., {Campana}, S., {Dall'Osso}, S., {Muno}, M.~P., {Cummings}, J.,
  {Perna}, R., \& {Stella}, L. 2007, \apj, 664, 448

\bibitem[{{Israel} {et~al.}(2012){Israel}, {Esposito}, {Rea}, {Turolla},
  {Zane}, {Mereghetti}, \& {Tiengo}}]{israel12}
{Israel}, G.~L., {Esposito}, P., {Rea}, N., {Turolla}, R., {Zane}, S.,
  {Mereghetti}, S., \& {Tiengo}, A. 2012, Astron. Tel., 3944

\bibitem[{{Jahoda} {et~al.}(1996){Jahoda}, {Swank}, {Giles}, {Stark},
  {Strohmayer}, {Zhang}, \& {Morgan}}]{jahoda96}
{Jahoda}, K., {Swank}, J.~H., {Giles}, A.~B., {Stark}, M.~J., {Strohmayer}, T.,
  {Zhang}, W., \& {Morgan}, E.~H. 1996, in SPIE Conference Series, Bellingham
  WA, Vol. 2808, EUV, X-Ray, and Gamma-Ray Instrumentation for Astronomy VII.,
  ed. O.~H.~W. {Siegmund} \& M.~A. {Gummin}, 59--70

\bibitem[{{Jansen} {et~al.}(2001){Jansen}, {Lumb}, {Altieri}, {Clavel}, {Ehle},
  {Erd}, {Gabriel}, {Guainazzi}, {Gondoin}, {Much}, {Munoz}, {Santos},
  {Schartel}, {Texier}, \& {Vacanti}}]{jansen01}
{Jansen}, F., et~al. 2001, \aap, 365,
  L1

\bibitem[{{Kaspi} {et~al.}(2003){Kaspi}, {Gavriil}, {Woods}, {Jensen},
  {Roberts}, \& {Chakrabarty}}]{kaspi03}
{Kaspi}, V.~M., {Gavriil}, F.~P., {Woods}, P.~M., {Jensen}, J.~B., {Roberts},
  M.~S.~E., \& {Chakrabarty}, D. 2003, \apjl, 588, L93

\bibitem[{{Koyama} {et~al.}(2007){Koyama}, {Tsunemi}, {Dotani}, {Bautz},
  {Hayashida}, {Tsuru}, {Matsumoto}, {Ogawara}, {Ricker}, {Doty}, {Kissel},
  {Foster}, {Nakajima}, {Yamaguchi}, {Mori}, {Sakano}, {Hamaguchi},
  {Nishiuchi}, {Miyata}, {Torii}, {Namiki}, {Katsuda}, {Matsuura}, {Miyauchi},
  {Anabuki}, {Tawa}, {Ozaki}, {Murakami}, {Maeda}, {Ichikawa}, {Prigozhin},
  {Boughan}, {Lamarr}, {Miller}, {Burke}, {Gregory}, {Pillsbury}, {Bamba},
  {Hiraga}, {Senda}, {Katayama}, {Kitamoto}, {Tsujimoto}, {Kohmura}, {Tsuboi},
  \& {Awaki}}]{koyama07}
{Koyama}, K., et~al. 2007, \pasj, 59, 23


\bibitem[{{Kouveliotou} {et~al.}(2003)}]{kouveliotou03} 
{Kouveliotou}, C., {Eichler}, D., {Woods}, P.~M., et al.\ 2003, \apjl, 596, L79 

\bibitem[{{Kuiper} \& {Hermsen}(2011)}]{kuiper11}
{Kuiper}, L. \& {Hermsen}, W. 2011, Astron. Tel., 3665

\bibitem[{{Lasker} {et~al.}(2008){Lasker}, {Lattanzi}, {McLean}, {Bucciarelli},
  {Drimmel}, {Garcia}, {Greene}, {Guglielmetti}, {Hanley}, {Hawkins},
  {Laidler}, {Loomis}, {Meakes}, {Mignani}, {Morbidelli}, {Morrison},
  {Pannunzio}, {Rosenberg}, {Sarasso}, {Smart}, {Spagna}, {Sturch},
  {Volpicelli}, {White}, {Wolfe}, \& {Zacchei}}]{lasker08}
{Lasker}, B.~M., et~al. 2008, \aj, 136, 735

\bibitem[{{Lawrence} {et~al.}(2007){Lawrence}, {Warren}, {Almaini}, {Edge},
  {Hambly}, {Jameson}, {Lucas}, {Casali}, {Adamson}, {Dye}, {Emerson},
  {Foucaud}, {Hewett}, {Hirst}, {Hodgkin}, {Irwin}, {Lodieu}, {McMahon},
  {Simpson}, {Smail}, {Mortlock}, \& {Folger}}]{lawrence07}
{Lawrence}, A., et~al. 2007, \mnras, 379, 1599

\bibitem[{{Livingstone} {et~al.}(2011){Livingstone}, {Scholz}, {Kaspi}, {Ng},
  \& {Gavriil}}]{livingstone11}
{Livingstone}, M.~A., {Scholz}, P., {Kaspi}, V.~M., {Ng}, C.-Y., \& {Gavriil},
  F.~P. 2011, \apjl, 743, L38

\bibitem[{{Lyubarsky}(2002)}]{lyubarsky02}
{Lyubarsky}, Y.~E. 2002, \mnras, 332, 199

\bibitem[{{Manchester} {et~al.}(2005){Manchester}, {Hobbs}, {Teoh}, \&
  {Hobbs}}]{manchester05}
{Manchester}, R.~N., {Hobbs}, G.~B., {Teoh}, A., \& {Hobbs}, M. 2005, \aj, 129,
  1993

\bibitem[{{Mereghetti}(2008)}]{mereghetti08}
{Mereghetti}, S. 2008, \aapr, 15, 225

\bibitem[{{Mignani}(2011)}]{mignani11}
{Mignani}, R.~P. 2011, Advances in Space Research, 47, 1281

\bibitem[{{Mitsuda} {et~al.}(2007){Mitsuda}, {Bautz}, {Inoue}, {Kelley},
  {Koyama}, {Kunieda}, {Makishima}, {Ogawara}, {Petre}, {Takahashi}, {Tsunemi},
  {White}, {Anabuki}, {Angelini}, {Arnaud}, {Awaki}, {Bamba}, {Boyce}, {Brown},
  {Chan}, {Cottam}, {Dotani}, {Doty}, {Ebisawa}, {Ezoe}, {Fabian}, {Figueroa},
  {Fujimoto}, {Fukazawa}, {Furusho}, {Furuzawa}, {Gendreau}, {Griffiths},
  {Haba}, {Hamaguchi}, {Harrus}, {Hasinger}, {Hatsukade}, {Hayashida}, {Henry},
  {Hiraga}, {Holt}, {Hornschemeier}, {Hughes}, {Hwang}, {Ishida}, {Ishisaki},
  {Isobe}, {Itoh}, {Iyomoto}, {Kahn}, {Kamae}, {Katagiri}, {Kataoka},
  {Katayama}, {Kawai}, {Kilbourne}, {Kinugasa}, {Kissel}, {Kitamoto}, {Kohama},
  {Kohmura}, {Kokubun}, {Kotani}, {Kotoku}, {Kubota}, {Madejski}, {Maeda},
  {Makino}, {Markowitz}, {Matsumoto}, {Matsumoto}, {Matsuoka}, {Matsushita},
  {McCammon}, {Mihara}, {Misaki}, {Miyata}, {Mizuno}, {Mori}, {Mori}, {Morii},
  {Moseley}, {Mukai}, {Murakami}, {Murakami}, {Mushotzky}, {Nagase}, {Namiki},
  {Negoro}, {Nakazawa}, {Nousek}, {Okajima}, {Ogasaka}, {Ohashi}, {Oshima},
  {Ota}, {Ozaki}, {Ozawa}, {Parmar}, {Pence}, {Porter}, {Reeves}, {Ricker},
  {Sakurai}, {Sanders}, {Senda}, {Serlemitsos}, {Shibata}, {Soong}, {Smith},
  {Suzuki}, {Szymkowiak}, {Takahashi}, {Tamagawa}, {Tamura}, {Tamura},
  {Tanaka}, {Tashiro}, {Tawara}, {Terada}, {Terashima}, {Tomida}, {Torii},
  {Tsuboi}, {Tsujimoto}, {Tsuru}, {Turner}, {Ueda}, {Ueno}, {Ueno}, {Uno},
  {Urata}, {Watanabe}, {Yamamoto}, {Yamaoka}, {Yamasaki}, {Yamashita},
  {Yamauchi}, {Yamauchi}, {Yaqoob}, {Yonetoku}, \& {Yoshida}}]{mitsuda07}
{Mitsuda}, K., et~al. 2007, \pasj, 59, 1

\bibitem[{{Nobili} {et~al.}(2008){Nobili}, {Turolla}, \& {Zane}}]{nobili08}
{Nobili}, L., {Turolla}, R., \& {Zane}, S. 2008, \mnras, 386, 1527

\bibitem[{{Pagani} {et~al.}(2011){Pagani}, {Beardmore}, \& {Kennea}}]{pagani11}
{Pagani}, C., {Beardmore}, A.~P., \& {Kennea}, J.~A. 2011, Astron. Tel., 3493


\bibitem[{{Perna} \& {Gotthelf} (2008)}]{pg08}
{Perna}, R. \& {Gotthelf}, E. V., 2008, \apj, 681, 522

\bibitem[{{Perna} \& {Pons} (2011)}]{perna11}
{Perna}, R. \& {Pons}, J.A., 2011, \apj, 727, L51

\bibitem[{{Pfeffermann} {et~al.}(1987){Pfeffermann}, {Briel}, {Hippmann},
  {Kettenring}, {Metzner}, {Predehl}, {Reger}, {Stephan}, {Zombeck},
  {Chappell}, \& {Murray}}]{pfeffermann87}
{Pfeffermann}, E., et~al. 1987, in SPIE Conference Series, Bellingham, WA, Vol.
  733, Soft X-ray optics and technology. Edited by E.-E.~Koch \& G.~Schmahl,
  519

\bibitem[{{Pons} {et~al.}(2009){Pons}, {Miralles}, \& {Geppert}}]{pons09}
{Pons}, J.~A., {Miralles}, J.~A., \& {Geppert}, U. 2009, \aap, 496, 207


\bibitem[{{Pons} \& {Perna} (2011)}]{pons11}
{Pons}, J.A. \& {Perna}, R. , 2011, \apj, 741, L123


\bibitem[{{Pons} \& {Rea}(2012)}]{pons12}
{Pons}, J.~A. \& {Rea}, N. 2012, \apj, 750, L6

\bibitem[{{Predehl} \& {Schmitt}(1995)}]{predehl95}
{Predehl}, P. \& {Schmitt}, J.~H.~M.~M. 1995, \aap, 293, 889

\bibitem[{Rea \& Esposito(2011)}]{rea11}
Rea, N. \& Esposito, P. 2011, in High-Energy Emission from Pulsars and their
  Systems, ed. D.~F. Torres \& N.~Rea, Astrophysics and Space Science
  Proceedings (Springer Berlin Heidelberg), 247--273

\bibitem[{{Rea} {et~al.}(2011{\natexlab{a}}){Rea}, {Esposito}, {Israel},
  {Tiengo}, \& {Zane}}]{rei11}
{Rea}, N., {Esposito}, P., {Israel}, G.~L., {Tiengo}, A., \& {Zane}, S.
  2011{\natexlab{a}}, Astron. Tel., 3501


\bibitem[{{Rea} {et~al.}(2010){Rea}, {Esposito}, {Turolla}, {Israel}, {Zane},
  {Stella}, {Mereghetti}, {Tiengo}, {G{\"o}tz}, {G{\"o}{\u g}{\"u}{\c s}}, \&
  {Kouveliotou}}]{rea10}
{Rea}, N., {Esposito}, P., {Turolla}, R., {Israel}, G.~L., {Zane}, S.,
  {Stella}, L., {Mereghetti}, S., {Tiengo}, A., {G{\"o}tz}, D., {G{\"o}{\u
  g}{\"u}{\c s}}, E., \& {Kouveliotou}, C. 2010, Science, 330, 944

\bibitem[{{Rea} {et~al.}(2011{\natexlab{b}}){Rea}, {Mignani}, {Israel}, \&
  {Esposito}}]{rmi11}
{Rea}, N., {Mignani}, R.~P., {Israel}, G.~L., \& {Esposito}, P.
  2011{\natexlab{b}}, Astron. Tel., 3515

\bibitem[{{Rea} {et~al.}(2012){Rea}, {Pons}, {Torres}, \& {Turolla}}]{rea12}
{Rea}, N., {Pons}, J.~A., {Torres}, D.~F., \& {Turolla}, R. 2012, \apjl, 748,
  L12

\bibitem[{{Rea} {et~al.}(2008){Rea}, {Zane}, {Turolla},{Lyutikov} \& {G\"otz}}]{rea08}
{Rea}, N., {Zane}, S., {Turolla}, R., {Lyutikov}, M. \& {G\"otz}, D. 2008, \apjl, 686, 1245


\bibitem[{{Rea} {et~al.}(2009){Rea}, {et al.}}]{rea09}
{Rea}, N., {et al.}  2009, \mnras, 396, 2419


\bibitem[{{Rea} {et~al.}(2007){Rea}, {Zane}, {Lyutikov} \& {Turolla}}]{rea07}
{Rea}, N., {Zane}, S., {Lyutikov}, M. \& {Turolla}, R. 2007, \apss, 308, 505



\bibitem[{{Skrutskie} {et~al.}(2006){Skrutskie}, {Cutri}, {Stiening},
  {Weinberg}, {Schneider}, {Carpenter}, {Beichman}, {Capps}, {Chester},
  {Elias}, {Huchra}, {Liebert}, {Lonsdale}, {Monet}, {Price}, {Seitzer},
  {Jarrett}, {Kirkpatrick}, {Gizis}, {Howard}, {Evans}, {Fowler}, {Fullmer},
  {Hurt}, {Light}, {Kopan}, {Marsh}, {McCallon}, {Tam}, {Van Dyk}, \&
  {Wheelock}}]{skrutskie06}
{Skrutskie}, M.~F., et~al. 2006, \aj, 131, 1163

\bibitem[{{Thompson}(2008)}]{thompson08}
{Thompson}, C. 2008, \apj, 688, 499

\bibitem[{{Thompson} {et~al.}(2002){Thompson}, {Lyutikov}, \&
  {Kulkarni}}]{tlk02}
{Thompson}, C., {Lyutikov}, M., \& {Kulkarni}, S.~R. 2002, \apj, 574, 332

\bibitem[{{Turolla}(2009)}]{turolla09}
{Turolla}, R. 2009, in Astrophysics and Space Science Library. Springer Berlin,
  Heidelberg, Vol. 357, Neutron stars and pulsars, ed. {W.~Becker}, 141

\bibitem[{{Turolla} {et~al.}(2011){Turolla}, {Zane}, {Pons}, {Esposito}, \&
  {Rea}}]{turolla11}
{Turolla}, R., {Zane}, S., {Pons}, J.~A., {Esposito}, P., \& {Rea}, N. 2011,
  \apj, 740, 105


\bibitem[{{van der Horst} {et al.}(2010)}]{vdh10}
{van der Horst}, A. J.,  {Connaughton}, V., {Kouveliotou}, C, et~al. 2010, \apj, 711, L1


\bibitem[{{van Straten} \& {Bailes}(2011)}]{vanstraten11}
{van Straten}, W. \& {Bailes}, M. 2011, PASA, 28, 1

\bibitem[{{Voges} {et~al.}(1999){Voges}, {Aschenbach}, {Boller},
  {Br{\"a}uninger}, {Briel}, {Burkert}, {Dennerl}, {Englhauser}, {Gruber},
  {Haberl}, {Hartner}, {Hasinger}, {K{\"u}rster}, {Pfeffermann}, {Pietsch},
  {Predehl}, {Rosso}, {Schmitt}, {Tr{\"u}mper}, \& {Zimmermann}}]{voges99}
{Voges}, W., et~al. 1999, \aap, 349, 389

\bibitem[{{Wang} {et~al.}(2008){Wang}, {Bassa}, {Kaspi}, {Bryant}, \&
  {Morrell}}]{wang08}
{Wang}, Z., {Bassa}, C., {Kaspi}, V.~M., {Bryant}, J.~J., \& {Morrell}, N.
  2008, \apj, 679, 1443

\bibitem[{{Woods} {et~al.}(2004){Woods}, {Kaspi}, {Thompson}, {Gavriil},
  {Marshall}, {Chakrabarty}, {Flanagan}, {Heyl}, \& {Hernquist}}]{woods04}
{Woods}, P.~M., {Kaspi}, V.~M., {Thompson}, C., {Gavriil}, F.~P., {Marshall},
  H.~L., {Chakrabarty}, D., {Flanagan}, K., {Heyl}, J., \& {Hernquist}, L.
  2004, \apj, 605, 378



\bibitem[{{Zane} {et~al.}(2009){Zane}, {Rea}, {Turolla}, \& {Nobili}}]{zane09}
{Zane}, S., {Rea}, N., {Turolla}, R., \& {Nobili}, L.. 2009, \mnras , 398, 1403

\bibitem[{{Zombeck} {et~al.}(1995}]{hrc95}
{Zombeck}, M. V., et al. 1995, SPIE, 2518, 96


\end{thebibliography}

\end{document}